\documentclass{optica-article}
\journal{oe}
\articletype{Research Article}

\usepackage{graphicx}
\usepackage{adjustbox}
\usepackage{orcidlink}
\usepackage[font=scriptsize, textfont=sf]{subcaption}
\usepackage{tabularray}
\usepackage{rotate}
\usepackage{ifthen}
\usepackage{url}
\usepackage[nopostdot,symbols,nogroupskip,nonumberlist]{glossaries-extra}
\definecolor{added}{HTML}{004D92}

\usepackage{empheq}
\usepackage{makecell}
\usepackage{bm}
\usepackage{siunitx}

\newcommand\chem[1]{\ensuremath{\mathrm{#1}}}
\newcommand{\T}{^{\textsf{T}}}
\newcommand{\range}[3]
{%
     \ifthenelse{\equal{#1}{}}{
     	\left[#2,\,...\,,\,#3\right]
     }{%
     	#1\in\left[#2,\,...\,,\,#3\right]
     }%
}
\glssetcategoryattribute{acronym}{indexonlyfirst}{true}
\newignoredglossary{ignored}

\newabbreviation{hs}{HS}{hyperspectral}
\newabbreviation{rgb}{RGB}{red-green-blue}
\newabbreviation{swir}{SWIR}{short wave infrared}
\newabbreviation{nir}{NIR}{near infrared response}
\newabbreviation{vis}{VIS}{visible}
\newabbreviation{uv}{UV}{ultraviolet}
\newabbreviation{ir}{IR}{infrared}

\newabbreviation{gsd}{GSD}{ground sample distance}
\newabbreviation{snr}{SNR}{signal to noise ratio}
\newabbreviation{mtf}{MTF}{modulation transfer function}

\newabbreviation{psnr}{PSNR}{peak \glsentrylong{snr}}
\newabbreviation{scc}{sCC}{spatial cross-covariance coefficient}
\newabbreviation{ergas}{ERGAS}{relative dimensionless global error in synthesis}
\newabbreviation{sam}{SAM}{spectral angle mapper}
\newabbreviation{ssim}{SSIM}{structural similarity}

\newabbreviation{mse}{MSE}{mean square error}
\newabbreviation{rmse}{RMSE}{root \glsentrylong{mse}}
\newabbreviation{awgn}{AWGN}{additive white Gaussian noise}
\newabbreviation{std}{STD}{standard deviation}
\newabbreviation{dft}{DFT}{discrete Fourier transform}

\newabbreviation{ml}{ML}{maximum likelihood}
\newabbreviation{mle}{MLE}{maximum likelihood estimation}
\newabbreviation{es}{ES}{exhaustive search}
\newabbreviation{gn}{TRR}{trust region refinement}
\newabbreviation{lm}{LM}{Levenberg-Marquardt}

\newabbreviation{fts}{FTS}{Fourier transform spectrometer}
\newabbreviation{ftir}{FTIR}{Fourier transform infrared spectroscopy}
\newabbreviation{fpa}{FPA}{focal plane array}
\newabbreviation{fov}{FoV}{field of view}
\newabbreviation{fp}{FP}{Fabry-Perot}
\newabbreviation{opd}{OPD}{optical path difference}
\newabbreviation{opl}{OPL}{optical path length}
\newabbreviation{ft}{FT}{Fourier transform}

\newabbreviation[longplural={greenhouse gases}]{ghg}{GHG}{greenhouse gas}
\newabbreviation{ehs}{EHS}{environment, health and safety}

\newabbreviation{irca}{IRCA}{interferometer response characterization algorithm}
\newabbreviation{fui}{FUI}{Fonds Unique Interministériel}
\newabbreviation{anr}{ANR}{Agence Nationale de Recherche}
\newabbreviation{ipag}{IPAG}{Institut de Planétologie et d'Astrophysique de Grenoble}
\newabbreviation{onera}{ONERA}{Office National d'Etudes et de Recherches Aérospatiales}
\newabbreviation{imspoc}{ImSPOC}{Image SPectrometer On Chip}
\newabbreviation{tralfic}{TRALFIC}{trust region algorithm for low finesse interferometer characterization}
\newabbreviation[type=ignored]{imagaz}{ImaGAZ}{ImaGAZ}
\newabbreviation{scarbo}{SCARBO}{Space CARBon Observatory}
\newabbreviation[type=ignored]{imspoc-uv}{ImSPOC-UV}{\glsentryshort{imspoc}-\glsentrylong{uv}}
\newabbreviation{presto}{PRESTO}{Precursory Research for Embryonic Science and Technology}
\newabbreviation{nasa}{NASA}{National Aeronautics and Space Administration}
\newabbreviation{h2020}{H2020}{Horizon 2020}
\newabbreviation{fumultispoc}{FuMultiSPOC}{FUsion MULTIspectral-\glsentryshort{imspoc}}

\newabbreviation{p1}{prototype 1}{prototype \glsentryshort{imspoc-uv}/\glsentryshort{vis}}
\newabbreviation{p2}{prototype 2}{prototype \glsentryshort{imspoc-uv}-drone}
\newabbreviation{p3}{prototype 4}{prototype \glsentryshort{imagaz}-1}
\newabbreviation{p4}{PROTO-4}{prototype NanoCarb-1}
\newabbreviation{p5}{prototype 3}{prototype WFAI}
\graphicspath{{figures}}

\begin{document}

\title{%
	Interferometer response characterization algorithm for multi-aperture Fabry-Perot imaging spectrometers
}

\author{
	Daniele Picone,\authormark{1,2}\,\orcidlink{0000-0002-0226-8399}
	Silv\`{e}re Gousset,\authormark{2}\,\orcidlink{0000-0003-3106-8262}
	Mauro Dalla Mura,\authormark{1,3,*}\,\orcidlink{0000-0002-9656-9087}
	Yann Ferrec,\authormark{4}\,\orcidlink{0000-0002-8361-5398}
	and
	Etienne le Coarer\authormark{2}\,\orcidlink{0000-0001-6571-5494}
}

\address{
	\authormark{1} Univ. Grenoble Alpes, CNRS, Grenoble INP \authormark{$\dagger$}, GIPSA-lab,
	38000 Grenoble, France\\
	\authormark{2} Univ. Grenoble Alpes, CNRS, IPAG,
	38000 Grenoble, France\\
	\authormark{3} Institut Universitaire de France (IUF), 75005 Paris, France\\
	\authormark{4} ONERA/DOTA, BP 80100, chemin de la Huni\`{e}re, 91123 Palaiseau, France\\
	\authormark{$\dagger$} Institute of Engineering Univ. Grenoble Alpes
}
\email{\authormark{*}mauro.dalla-mura@grenoble-inp.fr}

\begin{abstract}
	In recent years, the demand for hyperspectral imaging devices has grown significantly, driven by their ability of capturing high-resolution spectral information.
	Among the several possible optical designs for acquiring hyperspectral images, there is a growing interest in interferometric spectral imaging systems based on division of aperture. These systems have the advantage of capturing snapshot acquisitions while maintaining a compact design. However, they require a careful calibration to operate properly.
	In this work, we present the interferometer response characterization algorithm (IRCA), a robust three-step procedure designed to characterize the transmittance response of multi-aperture imaging spectrometers based on the interferometry of Fabry-Perot.
	Additionally, we propose a formulation of the image formation model for such devices suitable to estimate the parameters of interest by considering the model under various regimes of finesse.
	The proposed algorithm processes the image output obtained from a set of monochromatic light sources and refines the results using nonlinear regression after an ad-hoc initialization. Through experimental analysis conducted on four different prototypes from the Image SPectrometer On Chip (ImSPOC) family, we validate the performance of our approach for characterization. The associated source code for this paper is available at \url{https://github.com/danaroth83/irca}.
\end{abstract}

\glsresetall

\section{Introduction}
\label{sec:intro} 
    
    The demand for imaging spectrometers, also known as \gls{hs} cameras, has experienced significant growth in recent years. This surge in popularity can be attributed to their outstanding ability to capture high-resolution spectral information, especially in comparison to classic multispectral devices.
    These cameras find applications in various fields, such as astronomy, precision agriculture, molecular biology, biomedical imaging, geosciences, physics, and surveillance~\cite{Till07, BenD09, Adam10, Kim17}. Of particular importance is their role in accurately measuring atmospheric gases, which is vital for climate change monitoring, air quality studies, and compliance to regulatory requirements~\cite{Gous19}.
    
    Traditional imaging spectrometers that rely on scanning mechanism, such as whiskbroom and pushbroom, face limitations in capturing spatially varying scenes and are forced to make compromises between spectral and spatial resolution~\cite{Bors21}.
    Consequently, significant research efforts have been recently dedicated to the development and production of computational spectral imaging systems. These systems aim to enhance spectral, spatial, and temporal resolution and operate by encoding hyperspectral information in low-dimensional projected domains. However the retrieval of the full spectral and spatial \gls{hs} datacube requires the application of sophisticated reconstruction algorithms~\cite{Bacc23, Huan22}.

\begin{figure}
    \centering
    	
 	\begin{subfigure}[b]{.74\linewidth}
    	\centering
    	\includegraphics[width=\linewidth]{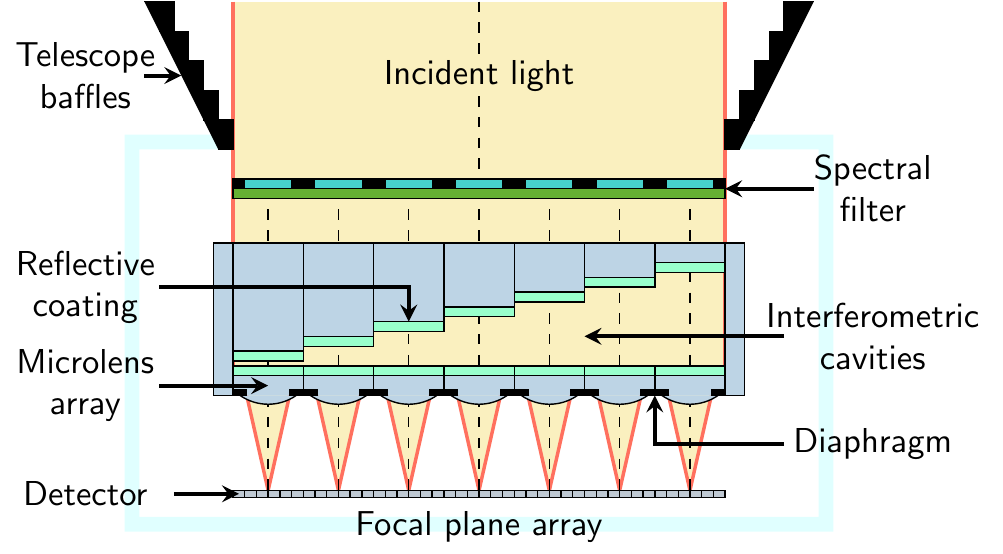}
    	\caption{Optical concept.}
    	\label{fig:concept_imspoc}
	\end{subfigure}
	\hfil
	\begin{minipage}[b]{.25\textwidth}
		
		\begin{subfigure}[b]{\linewidth}
			\centering
			\adjincludegraphics[width=\linewidth,trim={.2\width} {.0\height} {.15\width} {.1\height}, clip]{extra/imspoc_proto_imspoc_uv_2.jpg}
			\caption{\Glsentryshort{imspoc} \glsentryshort{p2}.}
			\label{fig:concept_imspoc_uv}
		\end{subfigure}
		
		\smallskip
		
	    \begin{subfigure}[b]{\linewidth}
			\centering
			\scalebox{-1}[1]{\adjincludegraphics[width=\linewidth,trim={.4\width} {.15\height} {.2\width} {.45\height}, clip]{extra/imspoc_proto_wfai_1.jpg}}
			\caption{\Glsentryshort{imspoc} \glsentryshort{p5}.}
			\label{fig:concept_wfai}
		\end{subfigure}
	
	\end{minipage}
    \caption[\Glsentryshort{imspoc} concept]{
    	Optical concept of multi-aperture interferometric imaging spectrometers and examples of \glsentryshort{imspoc} devices.
		(\subref{fig:concept_imspoc}) Cross-section view of the optical concept for an \glsentryshort{imspoc} imaging system. In the pictured design, the interferometers are air cavities of different thickness carved within a glass optical plate and coated with a reflective layer in titanium dioxide (\chem{TiO_2}). An external spectral filter can be also used to limit the input wavenumber range of the device for certain applications, such as \chem{NO_2} gas detection~\cite{Dole21}.
    	(\subref{fig:concept_imspoc_uv}), (\subref{fig:concept_wfai}) Examples of \gls{imspoc} prototypes; their specific characteristics are described in the experimental section.
    }
    \label{fig:concept}
\end{figure}

   	In this paper, we focus our attention on the characterization of the transmittance response of multi-aperture interferometric imaging spectrometers~\cite{Oikn18}. This class of instruments includes miniaturized snapshot acquisition systems for \gls{hs} imagery, whose optical design consists of a matrix of microlenses and a staircase-shaped optical plate superposed to a focal plane array. 
   	\figurename~\ref{fig:concept} shows the optical design of one of such devices, known as \gls{imspoc}~\cite{Ferr19,Gous19,Gous18,Gous17,Guer18}.

\begin{figure*}
       	\centering
       	\includegraphics[width=\linewidth]{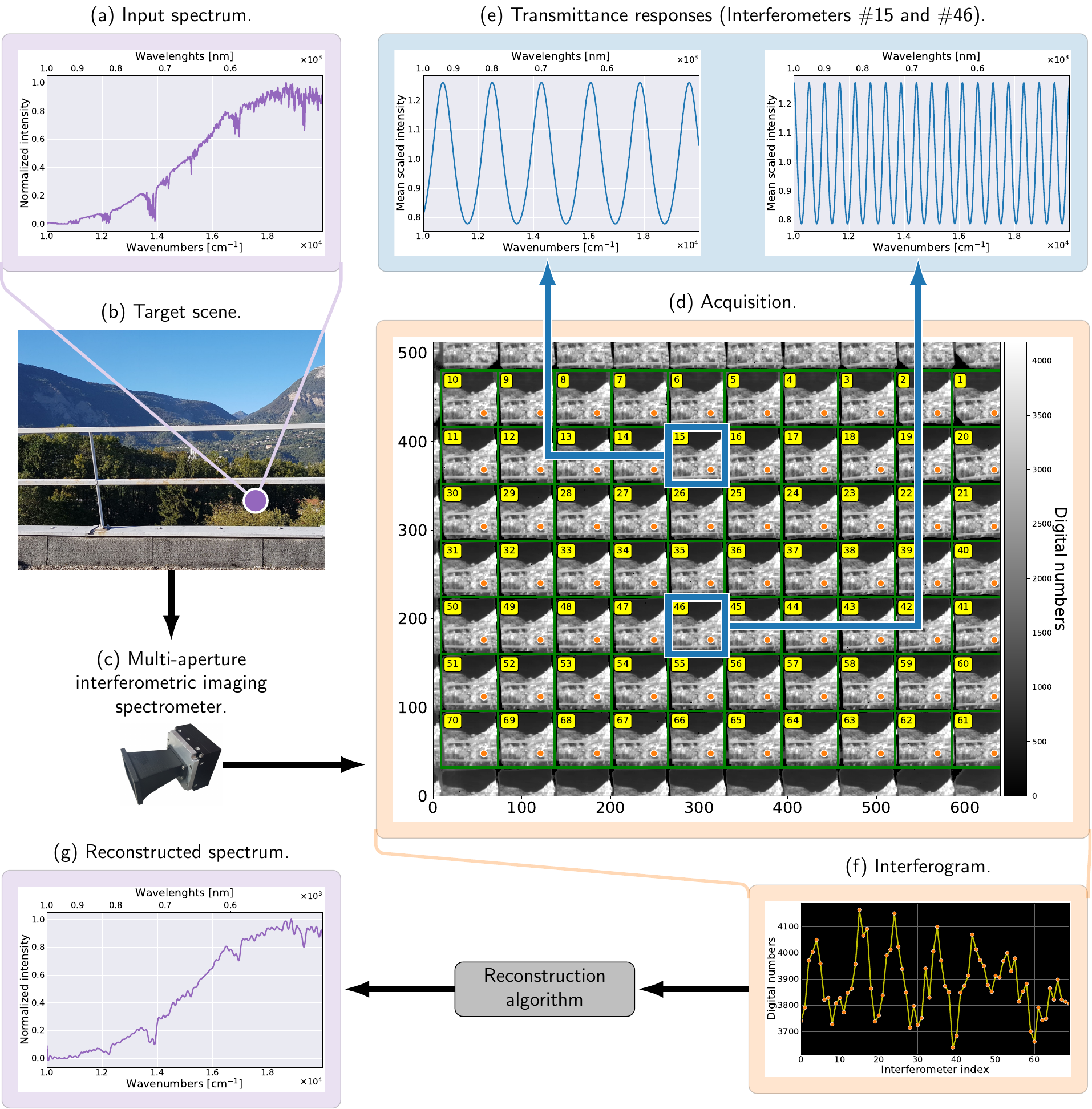}
        \caption[\Glsentryshort{imspoc} example acquisition]{
        	Visual representation of the operating principle of a multi-aperture interferometric imaging spectrometer.
        	In this example, the acquisition is a $512 \times 640$ image taken by an \glsentryshort{imspoc} device. The image is filtered by 70 interferometer/microlens units, resulting in the scene being split into multiple subimages. Matching pixels across these subimages are then arranged to form an interferogram for each ground point, which is subsequently processed to reconstruct the spectrum~\cite{Joun23, Pico21}. The purpose of this figure is purely illustrative.
        }
        \label{fig:example}
    \end{figure*}

   	\figurename~\ref{fig:example} illustrates an example of acquisition. The acquired image is composed of several subimages, with each subimage being the result of filtering the incident radiance using the transmittance response of a unique interferometer/microlens unit.
   	The set of readings obtained from the various subimages, arranged in ascending order of interferometer thicknesses, can be viewed as a sampled representation of a continuous interferogram linked to the spectrum of the particular region of the scene being viewed by the device.
   	In comparison to traditional hyperspectral cameras, multi-aperture interferometric imaging spectrometers offer several advantages such as snapshot acquisitions, compact dimensions, while preserving competitive spectral resolutions ranging from 5 to 10 \si{nm}. However, they do face limitations in terms of spatial resolution and potential parallax effects.

\begin{figure*}
        \centering
    	\includegraphics[width=\linewidth]{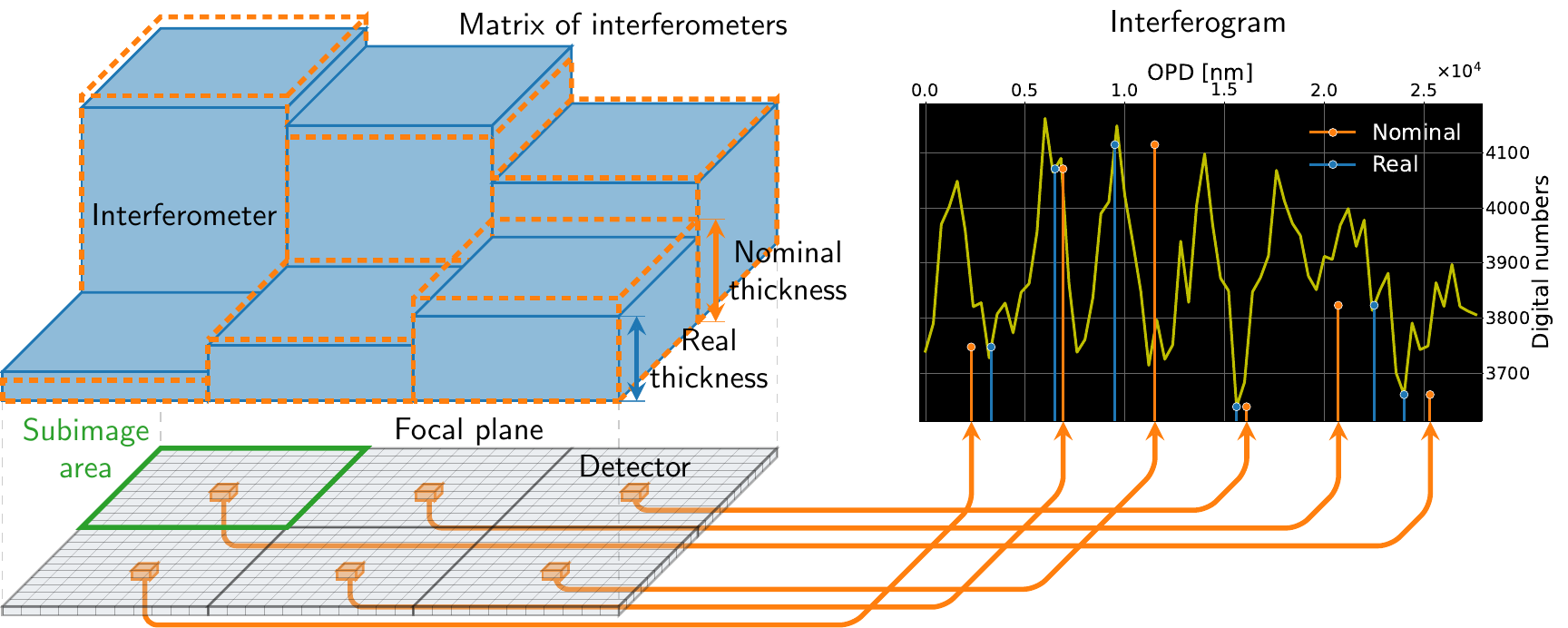}
        \caption[\Glsentryshort{opd} deviation effect on interferogram sampling]{	
        	Illustration of the effect of the errors in the estimation of the interferometer thickness. If the thicknesses of the optical components are known with an error (depicted by an orange dashed outline), the interferometer may sample the interferogram at a position that does not accurately correspond to the actual \glsentryshort{opd} value. In the specific case of the presented interferogram, the samples shift from the blue to the orange positions.	
        }
        \label{fig:opd_shift}
    \end{figure*}

    The identification of the image formation model for these devices is a crucial step that boils down to the estimation of the instrument optical transformation.
    Furthermore, regular calibration of the instrument becomes essential to maintain up-to-date device characterizations. This is especially relevant when considering potential changes in the instrument's physical properties over time, resulting from factors like instrument aging or variations in acquisition conditions, such as temperature.
    As an example scenario, due to the limited accuracy in either manufacturing or assembling the various device parts, the real thickness of the interferometers may be different with respect to the value they were designed for. As shown in \figurename~\ref{fig:opd_shift}, if this information is not taken into account, the interferogram samples are then placed incorrectly in domain of the \gls{opd}, which may cause inaccuracies on the quality of the reconstructed spectrum.
	
   	To address this issue, we propose in this work a general procedure for the parametric characterization of the instrument. The procedure is divided into a measurement session, where the device is illuminated with a set of flat field monochromatic sources, and an algorithmic estimation of the parameters of interest for the transmittance response, which we coin as \gls{irca}.
   	
   	The image formation model for the devices that we aim to characterize is also recalled in this work.
   	However, with respect to the typical formulation of the literature, we derive the formation model in terms of the parameters of interest. We formalize the mathematical model of the transmittance response of the device, specifically in terms of the \gls{opd}, reflectivity, gain, and phase shift.
   	We also express this transmittance response under different finesse regimes. Each regime corresponds to a specific number of emerging waves in the \gls{fp} cavity, arranged in descending order of optical power.
   	Previous studies~\cite{Dole19, Pico21} implicitly characterized such devices under the assumption of 2 emerging waves, prioritizing conceptual simplicity over a precise parameter estimation.
   	Our formulation enables us to describe previous techniques within our proposed framework, allowing for the application of similar trade-offs if desired. 
   	
   	Other than for \gls{imspoc} devices, the \gls{irca} can also be potentially employed to characterize and regularly update the calibration of various devices that exhibit a response based on the interferometry of \glsentrylong{fp}. This includes compressive imagers~\cite{Oikn18, Oikn18a} and hyperspectral imaging systems with dielectric mirrors~\cite{Pisa09, Zucc14}, among others.
   	
   	The \gls{irca} is defined by a three step procedure: the overall optical gain is firstly addressed discarding any interferometric effect, then a first rough assessment of the remaining parameters is performed by casting the problem as a \gls{ml} estimation of the characteristics of a sinusoidal signal. We refine this estimation by casting the problem as a nonlinear regression and solving it with the \gls{lm} algorithm~\cite{More78}. 
   	The nonlinear regression approach was also employed in other works~\cite{Hasa18}, but we focus our attention here on a robust solution for optical devices whose sensors are particularly sensitive to noise, as the different parameters are made separable by imposing that their polynomial expression in terms of wavelength has a limited degree.
   	
   	To summarize, the novel contributions of this work are:
   	\begin{enumerate}
   		\item The formalization of the image formation principle of multi-aperture Fabry-Perot imaging spectrometers (interferometers, lenslet, etc.).
   		We define within a single framework the dependency on its characteristic parameters (\gls{opd}, gain, reflectivity, phase shift) and the regimes of finesse associated to different amounts of transmitted waves;

   		\item The development of the \gls{irca}, a procedure for the estimation of parameters for transmittance responses of devices operating as \gls{fp} interferometers;
   		
   		\item The definition of an experimental procedure for the characterization of multi-aperture interferometric imaging spectrometers, using monochromatic sources. We test the effectiveness of the proposed method on real acquisitions from four \gls{imspoc} prototypes with different characteristics.
   	\end{enumerate}
   	
   	The article is organized as follows: in Section~\ref{sec:formation} we describe the image formation model of the multi-aperture interferometric imaging spectrometers, in Section~\ref{sec:characterization} we describe the proposed spectral characterization setup and estimation algorithm, and in Section~\ref{sec:experiments} we evaluate its performances and discuss its results in relation to the physics of the devices.
	    
	\begin{table}[t]
	    \footnotesize
	    \NineColors{saturation=high}
	    \caption[Variables list]{Selection of variables used in this paper, grouped in their respective categories.}
	    \begin{adjustbox}{width=\linewidth}
	    \begin{tblr}
	    	{
	    		colspec = {cll|ll},
	    		vlines = {white},
	    		hlines = {white},
	    		vline{4} = {2}{-}{white},
	    		cell{1}{2-5} = {red3, fg=yellow9, font=\bfseries},
	    		cell{2-14}{1} = {red3, fg=yellow9, font=\bfseries},
	    		cell{3,5,7,9,11,13,15}{2-5} = {red9},
	    	}
	    	&Symbol&Description&Symbol&Description\\
	    	\SetCell[r=4]{c}{\rotatebox{90}{Acq. model}}&
	    	$\sigma$ & Wavenumbers &$\mathbf{s}$&Focal plane coordinates\\
	    	&$\bm{\omega}=(\theta^{[i]},\phi^{[i]})$& Direction of incidence &$\{\Omega_j\}_{\range{j}{1}{N_p}}$& Solid angle of incidence\\
	   		&$\mathcal{L}(\sigma,\bm{\omega})$ & Input spectral radiance&$\{\Phi_{jk}\}_{\range{j}{1}{N_p},\range{k}{1}{N_i}}$&Received flux\\
	   		&$\{S_k\}_{\range{k}{1}{N_i}}$&Entrance pupil surface&$\{d_k\}_{\range{k}{1}{N_i}}$&Interferometer thickness\\
	    	\SetCell[r=4]{c}{\rotatebox{90}{Parameters}}&
	    	$\bm{\delta}=\{\delta_i\}_{\range{i}{1}{N_i}}$& \glsentryshortpl{opd} & $\varphi^{ }_0$&Phase shift\\
	    	&$\mathcal{A}(\sigma)$&Gain &$\mathbf{a}=\{a_m\}_{\range{m}{0}{N_d}}$& Gain coefficients\\
	    	&$\mathcal{R}(\sigma)$&Surface reflectivity &$\mathbf{r}=\{r_m\}_{\range{m}{0}{N_d}}$& Reflectivity coefficients\\
	    	&$\bm{\beta}=\{\beta_m\}_{\range{m}{1}{N_m}}$& Vector of parameters &$\hat{\bm{\beta}}=\{\hat{\beta}_m\}_{\range{m}{1}{N_m}}$& Estimated parameters \\
	    	\SetCell[r=3]{c}{\rotatebox{90}{Acq. vectors}} &$\bm{\sigma}=\{\sigma_i\}_{\range{i}{1}{N_a}}$& Central wavenumbers& $T_{\bm{\beta}}(\sigma_i)=\{t_i\}_{\range{i}{1}{N_a}}$ & Transmittance response\\
	    	&$\mathbf{y}=\{y_i\}_{\range{i}{1}{N_a}}$ & Single pixel acquisition& $\mathbf{w}=\{w_i\}_{\range{i}{1}{N_a}}$ & Flat field pixel statistic \\
	    	& $\mathbf{u}=\{u_i\}_{\range{i}{1}{N_a}}$ & Neighborhood mean&$\mathbf{v}=\{v_i\}_{\range{i}{1}{N_a}}$ & Scaled neighborhood mean\\
	    	\SetCell[r=3]{c}{\rotatebox{90}{Amount}} &$N_a$ & Acquisitions & $N_i$ & Interferometers\\
	    	& $N_p$ & Pixels per interferometer & $W$ & Waves\\
	    	& $N_d$ & Degree & $N_m$ & Parameters\\
	 	\end{tblr}
	 	\end{adjustbox}
	    \label{tab:variables}
	\end{table}

\section{Image formation for multi-aperture interferometric imaging spectrometers}
\label{sec:formation}

	In this section, we describe the image formation model of a multi-aperture interferometric imaging spectrometer. We begin by deriving the expression of the transmittance response for a single \gls{fp} interferometer/microlens unit in Section~\ref{ssec:formation_optical}. We then specify it within our framework for different regimes of finesse in Section~\ref{ssec:formation_fabry_perot}. Finally, we identify the parameters of interest for their characterization with their respective model in Section~\ref{ssec:formation_characterization}.
	
	Following the literature of \glsentrylongpl{fts}, in this work the spectra and transmittance responses are expressed in terms of \textit{wavenumbers} $\sigma$, that is as the reciprocal of the wavelengths (e.g., a wavelength of 500 \si{nm} corresponds to $\sigma=2\times 10^4$ \si{cm^{-1}}), but the relevant plots include both wavenumber and wavelength scales. Furthermore, the vertical ordinates are appropriately labeled as \textit{normalized intensity} when the intensity is scaled by its maximum value, and as \textit{mean scaled intensity} when scaled by its mean value. In situations involving multiple plots, all plots are consistently scaled using the mean of the reference.
	For the reader's convenience, the variables used in this paper are shown in \tablename~\ref{tab:variables}, separated into variables for the continuous image formation model, for its parameters, for the acquisition vectors and the vector sizes. These variables will be formerly introduced when relevant to the discussion.

	\subsection{Optical transfer model}
	\label{ssec:formation_optical}
	
\begin{figure*}
			
			\begin{subfigure}[b]{.58\linewidth}
				\centering
				\includegraphics[width=\linewidth]{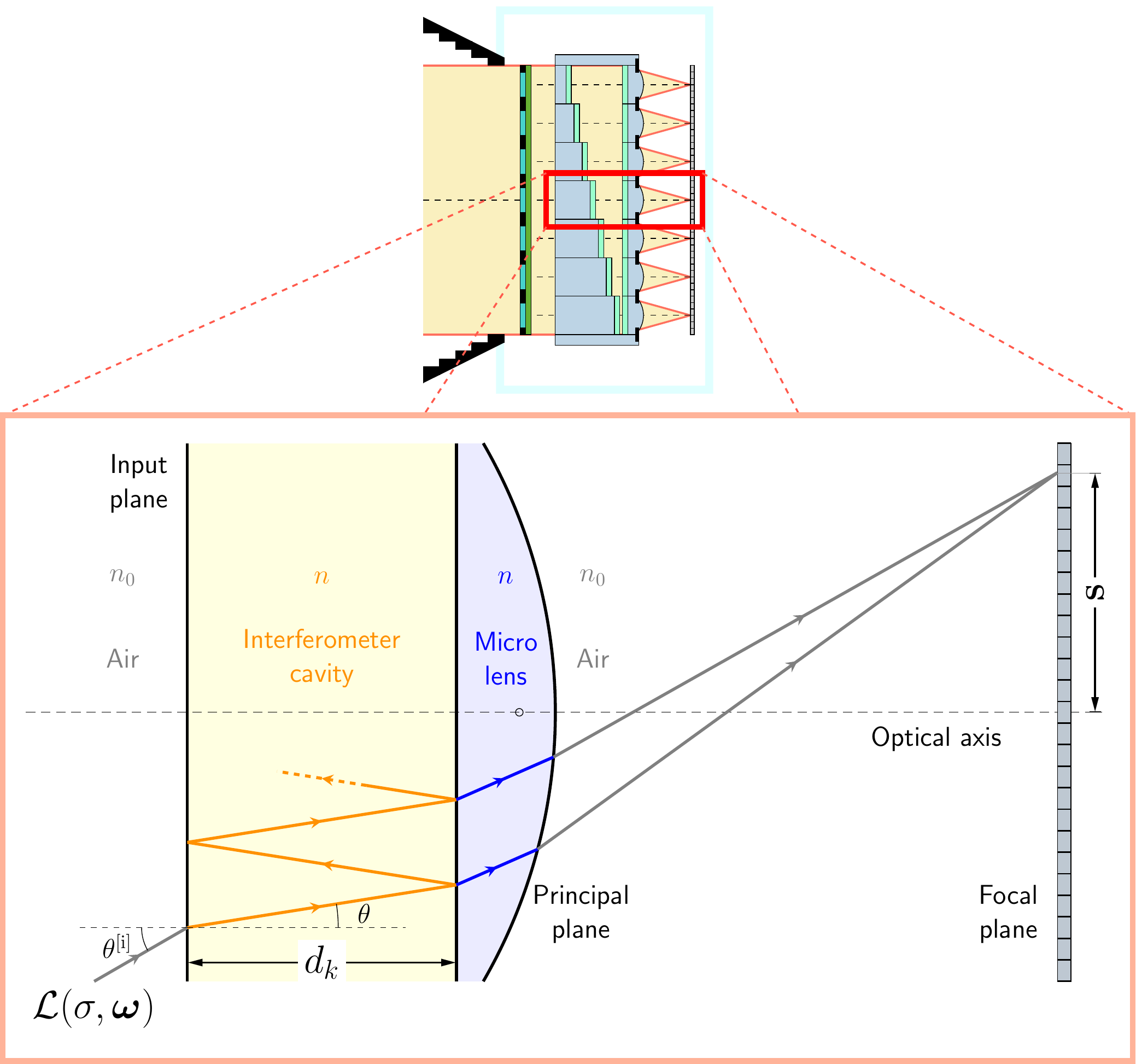}
				\caption{
					Focusing effect of a single \glsentryshort{fp} cavity/microlens unit.
				}
				\label{fig:formation_focusing}
			\end{subfigure}
			\hfil
			\begin{subfigure}[b]{.36\linewidth}
				\centering
				\includegraphics[width=\linewidth]{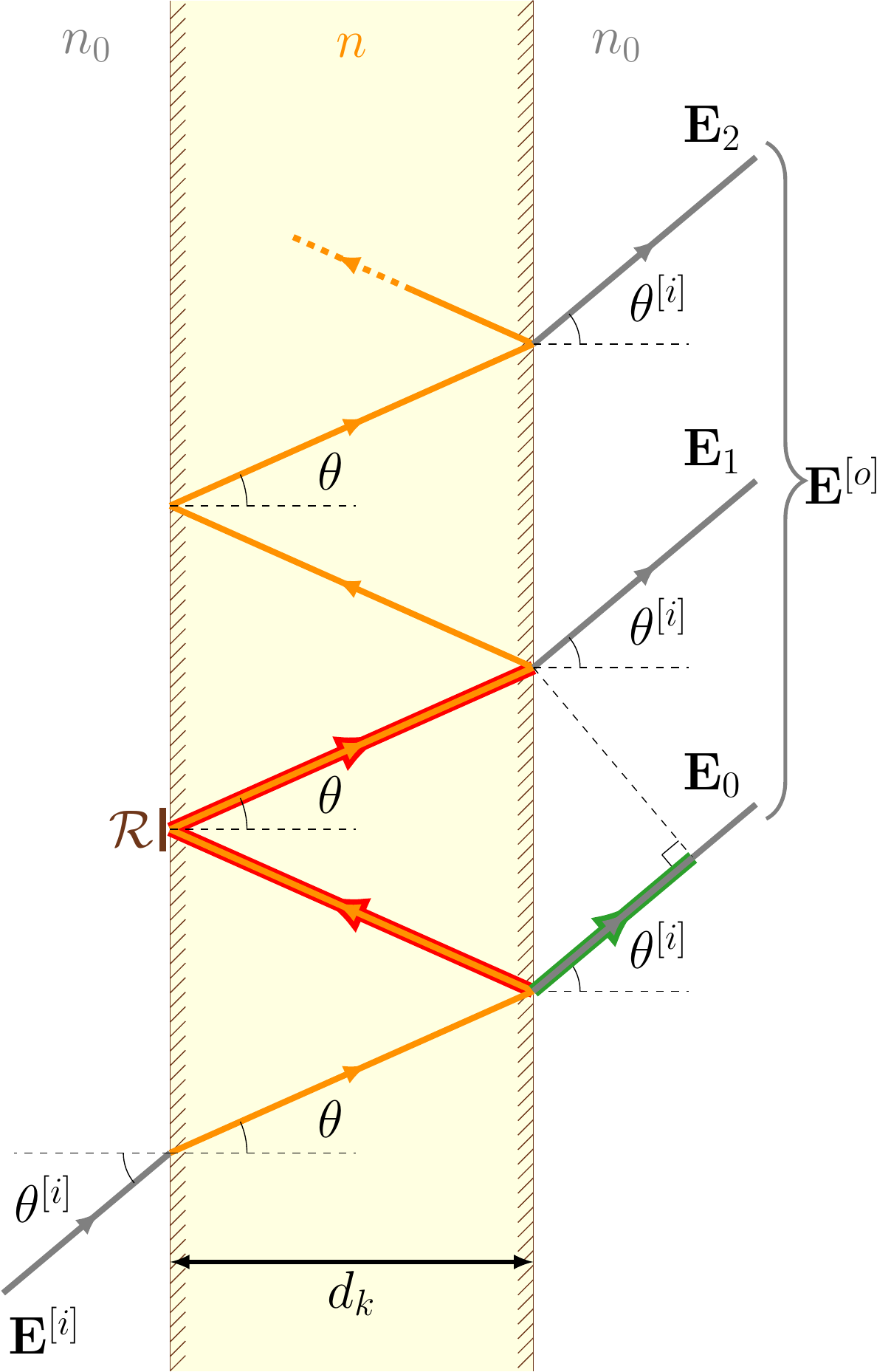}
				\caption{\Glsentrylong{fp} light ray propagation.}
				\label{fig:formation_fabry_perot}
			\end{subfigure}
		    \caption[\Glsentryshort{imspoc} image formation]{
				Light ray propagation within a multi-aperture \glsentryshort{fp} imaging spectrometer. (\subref{fig:formation_focusing}) Zoomed-in visualization of a single interferometer/microlens unit, showcasing the focusing effect of the light rays to the focal plane. (\subref{fig:formation_fabry_perot}) Detailed view of the \Glsentrylong{fp} cavity. The light propagation gives rise to an \glsentryshort{opd}, which corresponds to the difference between the optical paths highlighted in red and green.
		    }
		    \label{fig:formation}
		\end{figure*}

		We want to define here the expression of the sensors' readout in terms of the incident radiance. To this purpose, we analyze the light ray propagation within a single interferometer/lens unit of the optical system, as shown in \figurename~\ref{fig:formation_focusing}.
		
		By considering a scene at the optical infinity, there is a bijective correspondence between the direction of incidence $\bm{\omega}$ of the incident light and the position $\mathbf{s}$ on the focal plane.
		Consequently, we can express the spectral radiance $\mathcal{L}$ of the incident light either as $\mathcal{L}(\sigma,\,\bm{\omega})$ or $\mathcal{L}(\sigma,\,\mathbf{s})$.
		In this scenario, the $k$-th interferometer acts as a spectral filter and introduces an attenuation $\mathcal{T}_k$ of the radiant flux which varies only with the angle of incidence $\bm{\omega}$ and the wavenumber $\sigma$. As in the previous case, this can also be expressed interchangeably as $\mathcal{T}_k(\sigma,\,\bm{\omega})$ or $\mathcal{T}_k(\sigma,\,\mathbf{s})$.

		Assuming no crosstalk in the formation of each subimage, the spectral flux $\Phi_{jk}(\sigma)$ received by the $j$-th sensor (i.e., a photodetector) is only due to incident light within a given $k$-th interferometer. 
		Its expression at the focal plane is given by:
		\begin{equation}
			\Phi_{jk}(\sigma) = \int \mathcal{T}_k\left(\sigma,\,\mathbf{s}\right)\mathcal{L}\left(\sigma,\,\mathbf{s}\right)\;d\mathcal{G}\,,
			\label{eq:flux}
		\end{equation}
		where $\mathcal{G}$ denotes the geometric etendue subtended by the surface of the $j$-th photodetector and the exit pupil associated to the $k$-th microlens.
		
		Considering that the etendue is conserved across the object and the image space, we can rewrite eq.~\eqref{eq:flux} at the input plane as:
		\begin{equation}
			\Phi_{jk}(\sigma) = {S_k} \iint\limits_{\Omega_j}\mathcal{T}_k\left(\sigma,\,\bm{\omega}\right) \mathcal{L}\left(\sigma,\,\bm{\omega}\right) n_0\cos\theta^{[\mathrm{i}]} \,d\bm{\omega}\,,
			\label{eq:flux_etendue}
		\end{equation}
		where $\Omega_j$ is the solid angle of incident rays that focus over the $j$-th sensor, $S_k$ is the surface of the entrance pupil associated to the $k$-th interferometer, while $\theta^{[\mathrm{i}]}$ is the polar angle of the direction of incidence $\bm{\omega}$.

		Finally, we model the intensity level $x_{jk}$ captured by the photodetector as:
		\begin{equation}
			x_{jk}=\Delta t\int\limits_{\sigma_{min}}^{\sigma_{max}}\Phi_{jk}(\sigma)\,\xi(\sigma)\,\eta_j(\sigma)\,d\sigma\,,
			\label{eq:intensity}
		\end{equation} 
		where $[\sigma_{min}, \sigma_{max}]$ is the bandwidth of the instrument, $\eta_j(\sigma)$ denotes the quantum efficiency of the $j$-th sensor, $\xi(\sigma)$ denotes the spectral response of the accessory elements of the optical system (entry filter, leading optics, etc.), and $\Delta t$ denotes the integration time.
	        
	\subsection{\texorpdfstring{\Glsentrylong{fp}}{Fabry-Perot} regimes of finesse}
	\label{ssec:formation_fabry_perot}
	
        We now focus our attention on expanding the term $\mathcal{T}_k(\sigma,\,\bm{\omega})$ from eq.~\eqref{eq:flux_etendue}.
        Let us consider a monochromatic plane wave with complex amplitude $\mathbf{E}^{[\mathrm{i}]}(\sigma)$ incident to the \gls{fp} interferometer, forming an angle $\theta^{[\mathrm{i}]}$ with the normal to the incident plane.
        The complex amplitude $\mathbf{E}^{[\mathrm{o}]}$  of the transmitted light can be seen as a sum of $W\rightarrow\infty$ successive emerging waves $\{\mathbf{E}_m\}_{\range{m}{0}{W-1}}$.
        
        Each emerging wave introduces a fixed round trip phase difference: 
        \begin{equation}
        	\varphi=2\pi\delta\sigma-\varphi_0^{}\,,
        \end{equation}
    	where $\varphi_0^{}$ defines a constant phase shift and $\delta$ defines the \gls{opd} between two consecutive emerging waves.
        By referring to the geometry shown in \figurename~\ref{fig:formation_fabry_perot}, the \gls{opd} is determined as the difference between the optical paths for a round trip inner reflection and a direct transmission. 
        By making use of Snell's law, simple geometrical manipulations yield:
        \begin{equation}
        	\delta=
        	n\frac{2d_k}{\cos\theta}-n^{}_0(2d_k\tan\theta\sin\theta^{[\mathrm{i}]})=
        	n\left(\frac{2d_k}{\cos\theta}-2d_k\tan\theta\sin\theta\right)=
        	2nd_k\cos\theta\,,
        	\label{eq:opd_fabry_perot}
        \end{equation}
        where $d_k$ denotes the thickness of the $k$-th \gls{fp} cavity, while $n$ and $\theta$ are the refractive index and the reflection angle within the cavity, respectively.
        
       	In the following, we denote as $\mathcal{T}_{k}^{[W]}(\sigma, \bm{\omega})$ the expression of $\mathcal{T}_k(\sigma, \bm{\omega})$ specific to the a generic integer amount $W$ of emerging waves.
        Its expression is defined by the ratio between the output and input irradiance and evaluates as follows:
        
        \begin{subequations}
	        \begin{align}
	       		\mathcal{T}_{k}^{[W]}(\sigma, \bm{\omega})&:=\left\lvert\frac{\mathbf{E}^{[\mathrm{o}]}(\sigma)}{\mathbf{E}^{[\mathrm{i}]}(\sigma)}\right\rvert^2
	       		=(1-\mathcal{R})^2\left\lvert\sum_{m=0}^{W-1}\mathcal{R}^m\;\exp(-jm\varphi)\right\rvert^2\\
	       		&=(1-\mathcal{R})^2\left\lvert\frac{1-\mathcal{R}^{W}\exp(-jW\varphi)}{1-\mathcal{R}\exp(-j\varphi)}\right\rvert^2\,,
	        	\label{eq:airy_definition}
	        \end{align}
        \end{subequations}
        where $\mathcal{R}$ is the surface reflectivity, and the resulting term $(1-\mathcal{R})^2$ is due to the direct transmission through the cavity.
       	Specifically, for $2$, $W\ge0$ and $W\rightarrow\infty$ waves, we obtain:
       	\begin{subequations}
       		\begin{align}
       			\mathcal{T}_{k}^{[2]}(\sigma,\, \bm{\omega})&=\left(1+\mathcal{R}^2+2\mathcal{R}\cos\varphi\right)(1-\mathcal{R})^2\,,&\textrm{2 waves}\,,	\label{eq:airy_1}\\
       			\mathcal{T}_{k}^{[W]}(\sigma,\, \bm{\omega})&=\frac{1+\mathcal{R}^{2W}-2\mathcal{R}^{W}\cos(W\varphi)}{1+\mathcal{R}^2-2\mathcal{R}\cos\varphi}(1-\mathcal{R})^2\,,&\textrm{W waves}\,,\label{eq:airy_2}\\
       			\mathcal{T}_{k}^{[\infty]}(\sigma,\, \bm{\omega})&=\frac{(1-\mathcal{R})^2}{(1-\mathcal{R})^2+4\mathcal{R}\sin^2(\varphi/2)}\,,&\textrm{$\infty$ waves}\,.\label{eq:airy_3}
       		\end{align}
       		\label{eq:airy}
       	\end{subequations}
       	$\mathcal{T}_{k}^{[\infty]}(\sigma, \bm{\omega})$ is often known in the literature as the \textit{Airy distribution}~\cite{Isma16}.
       	 
       	For our purposes, it is also convenient to derive the mean scaled expression $\overline{\mathcal{T}}_{k}^{[W]}$ of $\mathcal{T}_{k}^{[W]}$ as:
       	\begin{equation}
       		\overline{\mathcal{T}}_{k}^{[W]}(\sigma,\bm{\omega})=\frac{1+\mathcal{R}}{(1-\mathcal{R}^{2W})(1-\mathcal{R})}\mathcal{T}_{k}^{[W]}(\sigma,\,\bm{\omega})\;.
       		\label{eq:norm}
       	\end{equation}
       	
       	\begin{figure*}
       		\begin{subfigure}[b]{0.47\linewidth}
       			\centering
       			\includegraphics[width=\linewidth]{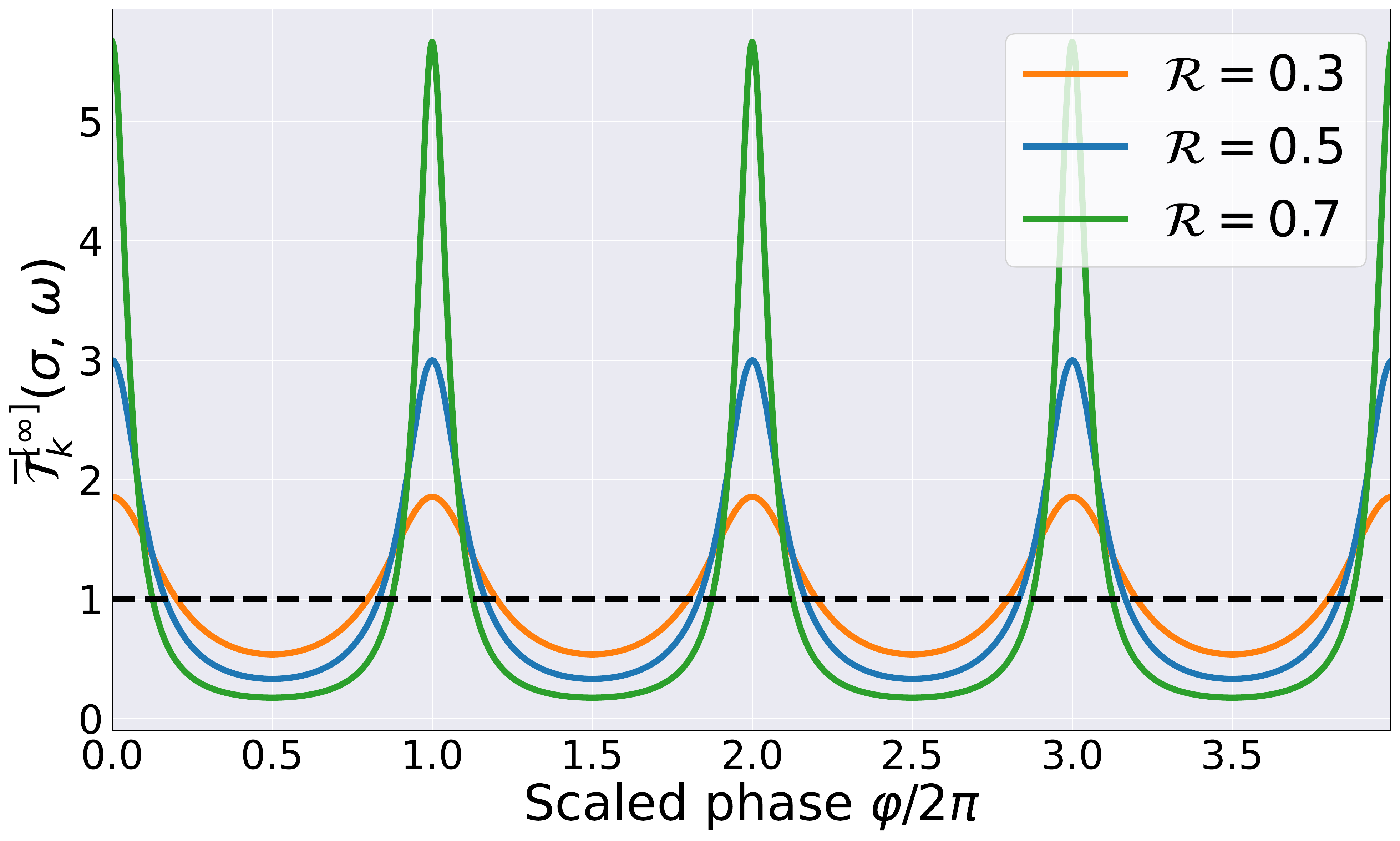}
       			\caption{Airy distribution.}
       			\label{fig:airy_finesse}
       		\end{subfigure}
       		\hfil
       		\begin{subfigure}[b]{0.5\linewidth}
       			\centering
       			\includegraphics[width=\linewidth]{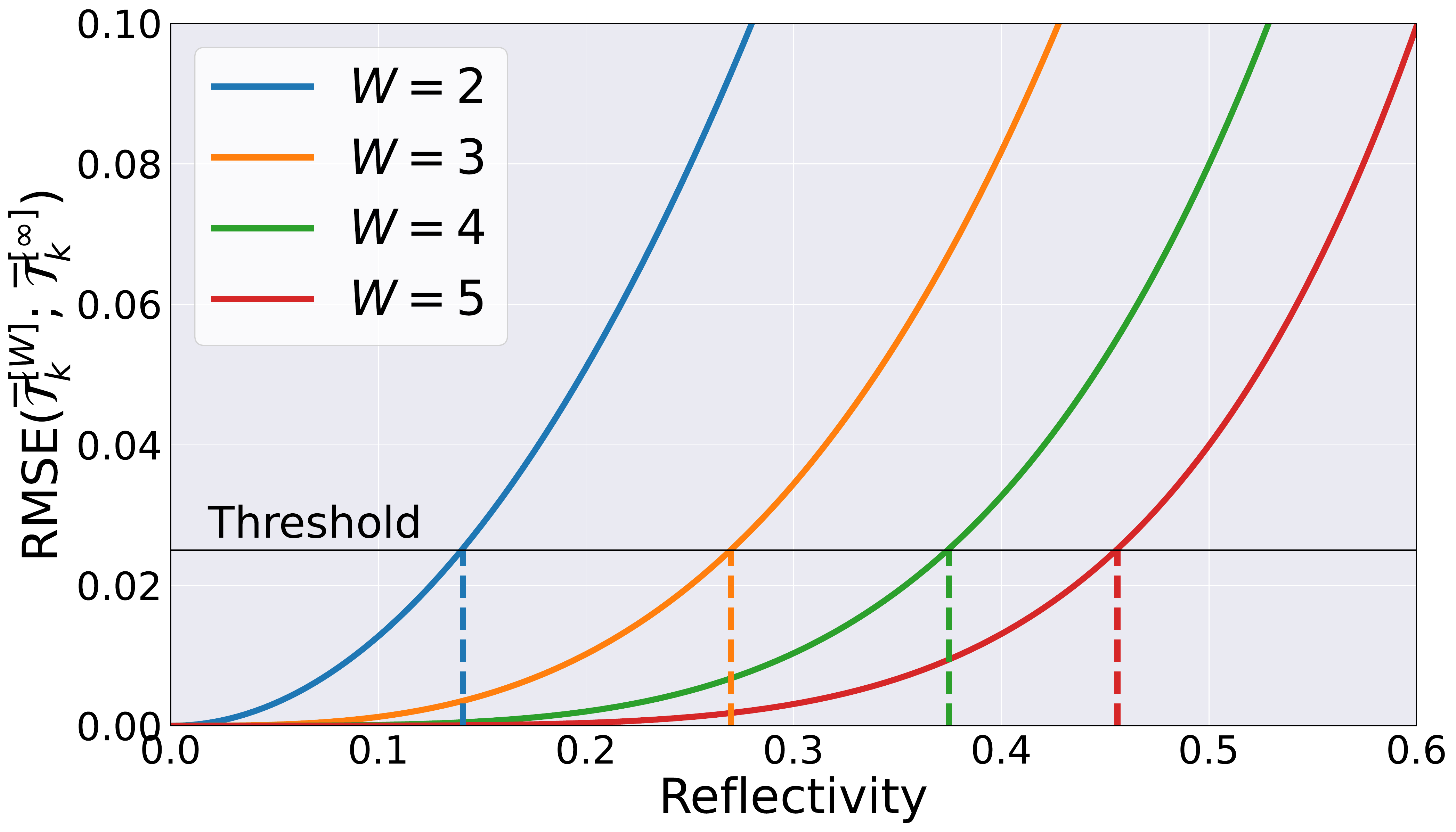}
       			\caption{Reflectivity regimes of $W$-wave models.}
       			\label{fig:airy_regimes}
       		\end{subfigure}
       		\caption[Airy distribution] {
       		Regimes of reflectivity. (\subref{fig:airy_finesse}) The figure illustrates the mean scaled theoretical transmittance response for various values of the surface reflectivity $\mathcal{R}$. (\subref{fig:airy_regimes})  Depending on the desired level of accuracy, the user can establish a \glsentrylong{rmse} threshold to identify the maximum reflectivity value (dashed line) in which the $W$-wave model of eq.~\eqref{eq:airy_2} exhibits behavior indistinguishable from the Airy distribution of eq.~\eqref{eq:airy_3}.
       		}
       		\label{fig:airy}
       	\end{figure*}

       	\figurename~\ref{fig:airy_finesse} presents the plot of the expression of $\overline{\mathcal{T}}_{k}^{[\infty]}$ for different values of reflectivity. This visualization assumes that $\mathcal{R}$ and $n$ remain constant regardless of the wavenumbers $\sigma$. However, this assumption is rarely verified in more realistic scenarios, where variations with respect to $\sigma$ are often observed.
       	The spacing between the peaks of the Airy distribution decreases as the thickness $d_k$ of the interferometer increases. This principle is exploited by multi-aperture devices to create different transmittance responses for different subimages as previously shown in \figurename~\ref{fig:example}.
       	       	
       	In the literature, transmittance responses are commonly classified based on the finesse parameter, whose value $4\mathcal{R}/(1-\mathcal{R})^2$ increases with the reflectivity.
        For low finesse devices, the response closely resembles a pure sinusoid. 
        This characteristic allows for increased throughput, resulting in higher \gls{snr} captured by the sensors. 
        In the case of high finesse, the spectral response exhibits sharper peaks, resulting in an enhanced periodic bandpass filtering effect.
 		Different finesse regimes can be determined for each wave model, by establishing the maximum reflectivity value such that a given error measure between the transmittance response of the $W$-wave model and the Airy distribution remains below a certain threshold. \figurename~\ref{fig:airy_regimes} illustrates this concept, employing the \gls{rmse} as the chosen error measure.
       	
       	\subsection{Proposed formulation of the image formation model}
       	\label{ssec:formation_characterization}
       	
       	In order to characterize the overall spectral response of the instrument at a given pixel, the physical acquisition model employed from eq.~\eqref{eq:flux_etendue} may be simplified, assuming that the optical transmittance response is roughly constant within the targeted solid angle $\Omega_j$ and $\cos\theta\approx 1$:
       	\begin{equation}
       		x_{jk}=\int\limits_{\sigma_{min}}^{\sigma_{max}}T_{\bm{\beta}}\left(\sigma\right)\left(\iint\limits_{\Omega_{j}} \mathcal{L}(\sigma,\,\bm{\omega})\,d\bm{\omega}\right)\,d\sigma\,.
       		\label{eq:intensity_simple}
       	\end{equation}
       	Here, $T_{\bm{\beta}}(\sigma)$ models the transmittance response of the entire instrument associated to a given pixel on the \gls{fpa}. For convenience, it is useful to describe it in terms of the expression of eq.~\eqref{eq:norm}:
       	\begin{equation}
       		T_{\bm{\beta}}(\sigma) = \mathcal{A}(\sigma)\overline{\mathcal{T}}_{k,W}(\sigma,\theta_j)
       		\label{eq:transfer_function}
       	\end{equation}
       	where we defined a gain variable:
       	\begin{equation} 					 		\mathcal{A}(\sigma):=\xi(\sigma)\eta_j(\sigma)S_k\Omega_j\frac{1+\mathcal{R}(\sigma)}{(1-\mathcal{R}^{2W}(\sigma))(1-\mathcal{R}(\sigma))}\;,
       		\label{eq:coupling}
       	\end{equation}
       	which incorporates all the multiplicative terms from eq.s~\eqref{eq:flux_etendue} and \eqref{eq:intensity}, while $\theta_j$ is the inner reflection angle associated to the incident light waves within the solid angle $\Omega_j$. The transmittance response is written in its scalar form so that the mean value with respect to $\sigma$ of $T_\beta(\sigma)$ is equal to that of $\mathcal{A}(\sigma)$.		

       	The terms $\mathcal{A}(\sigma)$ and $\mathcal{R}(\sigma)$ exhibit strong coupling in eq.~\eqref{eq:coupling}. To estimate their contributions separately, we impose them to be slowly varying functions with respect to $\sigma$. To achieve this, we restrict their models to polynomials of limited degree $N_d$:
       	\begin{equation}
       		\mathcal{A}(\sigma)=\sum_{m=0}^{N_d} a_m\sigma^m\,,\;\;\;\;\;\;\;\;\;\;\;\;\mathcal{R}(\sigma)=\sum_{m=0}^{N_d}r_m\sigma^m\,.\label{eq:polynomial}
       	\end{equation}
       	
       	The \gls{opd} value $\delta$ is assumed to be constant with the wavelength $\sigma$ as the rays interfer within the air in the prototypes under test (\figurename~\ref{fig:concept_imspoc}). This assumption is extended to the phase shift $\varphi^{}_0$ in order to simplify the computation.
       	These last hypotheses may be too limiting for interferometric cavities made of dispersive materials. For such cases, one may suppose a prior knowledge of this dispersion as function of $\sigma$ to reduce the problem to the estimation of the interferometer thickness, which is independent on $\sigma$.

       	Our goal then summarizes to find an estimation $\hat{\bm{\beta}}$ of the $2N_d+4$ elements of  $\bm{\beta}=\left[a_0\,,...,\,a_{N_d},\,r_0\,,...,\,r_{N_d},\,\delta,\,\varphi_0^{}\right]$ which allows to approximate the transmittance response $T_{\bm{\beta}}$ as accurately as possible.

\section{Proposed characterization procedure}
\label{sec:characterization}

	In this section, we present the proposed procedure for the spectral characterization of \gls{fp} interferometers. Specifically, we describe the measurement setup for the characterization in Section~\ref{ssec:characterization_setup} and we provide an overview of \gls{irca} in Section~\ref{ssec:characterization_overview}, detailing each of its composing steps in the subsequent sections.

	\subsection{Measurement setup for the characterization of the device}
	\label{ssec:characterization_setup}
	
		\begin{figure*}
		    \captionsetup[subfigure]{justification=centering}
		    \centering
		    \includegraphics[width=0.9\linewidth]{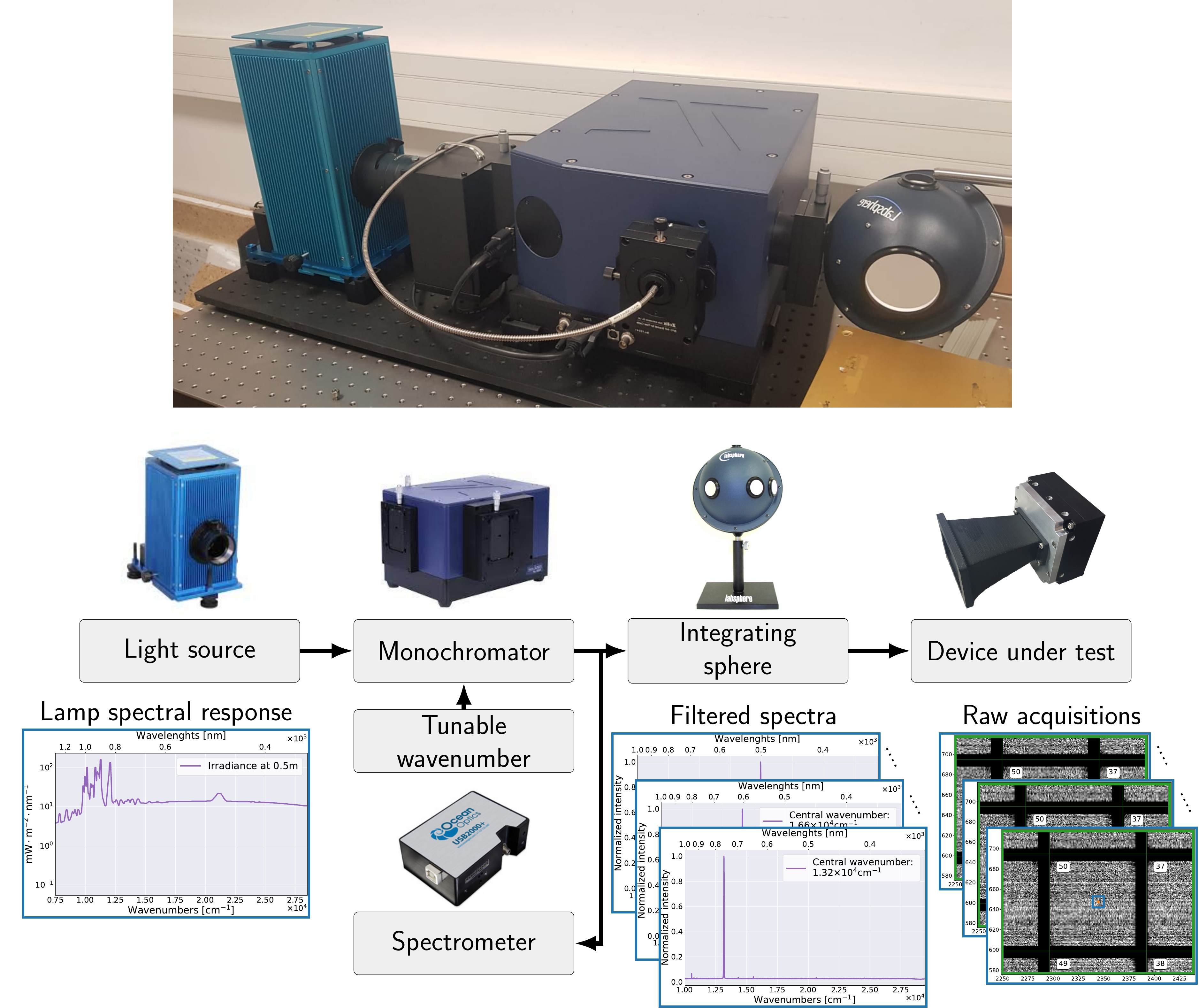}
		    \caption{Measurement setup for the characterization of multi-aperture interferometric imaging spectrometers.}
		    \label{fig:experimental_setup}
		\end{figure*}
		
		 To accurately characterize a given device under test and allow the inference of its parameters, it is necessary to capture a specific set of observations from reference sources under controlled conditions.
		 Perhaps the most straightforward approach involves illuminating the device with a flat field illumination using a set of monochromatic incident spectra with predefined central wavenumbers.
		 In fact, based on eq.~\eqref{eq:intensity_simple}, the corresponding $N_a$ acquisitions $\mathbf{y}\in\mathbb{R}^{N_a}$ are samples of the expected value of $T_{\bm{\beta}}(\sigma)$ evaluated at the specific wavenumbers $\bm{\sigma}\in\mathbb{R}^{N_a}$.
		 
		 In order to achieve this result, we propose the measurement setup shown in \figurename~\ref{fig:experimental_setup}. It involves the utilization of a wideband lamp as the light source, whose emitted light is filtered by a monochromator (i.e., equipped with a diffraction grating).
		 The bandwidth of the monochromator is deliberately narrower than the spectral resolution of the device, ensuring a sharply impulsive filtered spectrum.
		 Subsequently, this spectrum is uniformly scattered over the device under test by means of an integrating sphere.
		 By tuning the monochromator, a series of central wavenumbers is selected, and the device under test captures an image for each illumination in sequence. 
		 
		 An external spectrometer or probe is used to measure the incident power of the instrument. The measured value is used to equalize the energy of all the acquired images at different wavenumbers, with background level set to zero.
		 Finally, the vector $\mathbf{y}$ is obtained by extracting the specific spatial position from the acquired datacube, corresponding to the pixel being characterized. 
		 
		 Therefore, we formalize the problem at hand as finding the estimation $\hat{\bm{\beta}}$ of the parameter vector such that:
		 \begin{equation}
		 	\hat{\bm{\beta}}=\arg\min_{\bm{\beta}}\sum_{i=1}^{N_a} \left(T_{\bm{\beta}}(\sigma_i)-y_i\right)^2\;.
		 	\label{eq:cost}
		 \end{equation}
	
	\subsection{Overview of the \glsreset{irca}\gls{irca}}
	\label{ssec:characterization_overview}
	
		Solving eq.~\eqref{eq:cost} is a particularly challenging problem, due to the nonlinear dependency of $T_{\bm{\beta}}$ from the parameters $\bm{\beta}$. 
		The available tools for solving nonlinear regression methods are particularly sensitive to converging to non-local maxima~\cite{Rusz06}, so that a proper initialization is critical to produce an accurate parametrization of the optical system.
		
\begin{figure*}
	        \captionsetup[subfigure]{justification=centering}
	        \centering
	        \includegraphics[width=\linewidth]{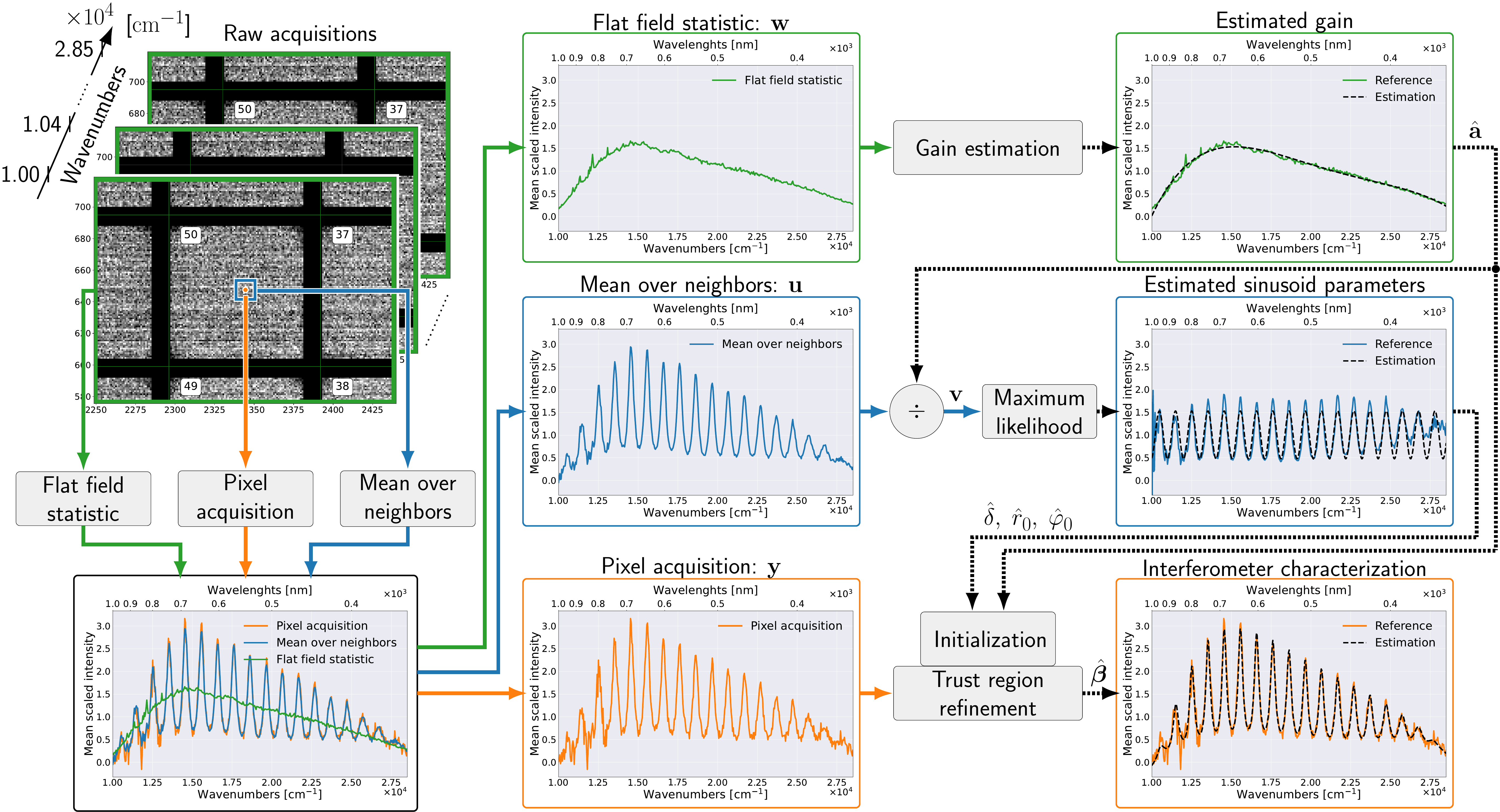}
	        \caption{Overview of the proposed \glsentryshort{irca} algorithm for the characterization of the transmittance response of a single \glsentrylong{fp} interferometer.}
	        \label{fig:algorithm}
	    \end{figure*}

		The proposed \glsreset{irca}\gls{irca}, depicted in \figurename~\ref{fig:algorithm}, consists of three steps, each dedicated to processing one of the three different sufficient statistics extracted from the $N_a$ images captured during the measurement session.
		This approach is designed to enhance the overall robustness of the final result by leveraging multiple aspects of the available information.
		We describe each of these steps below:
		\begin{itemize}
			\item The \textbf{gain estimation} step processes the vector $\mathbf{w}\in{\mathbb{R}^{N_a}}$, which represents the \textit{flat field statistic}. This vector is used to obtain an initial assessment of the gain coefficients  $\{\hat{a}_i\}_{\range{i}{0}{N_d}}$ of $\mathcal{A}(\sigma)$. 
			The flat field statistic captures the response of the pixel under test without the presence of the interferometric fringes. In cases where the vector $\mathbf{w}$ is not directly available, it can be approximated by evaluating the percentile from the raw acquisition across the entire focal plane. 
			This approach takes advantage of the global response of the image, which naturally dampens the oscillations caused by the interferometric fringes.
			\item The \textbf{\glsreset{ml}\gls{ml} initialization} involves processing the vector $\mathbf{u}\in\mathbb{R}^{N_a}$, which represents the \textit{mean over neighbors}.
			This step returns an initial estimation $\left[\hat{\delta},\,\hat{r}_0,\,\hat{\varphi}_0\right]$ of the remaining parameters, namely the \gls{opd}, reflectivity, and phase shift, respectively.
			The vector $\mathbf{u}$ is obtained by calculating the spatial average of the raw acquisition within a square window centered around the pixel under test. This averaging helps reduce the noise associated with the acquisition and could be performed even in the temporal domain, if such information is available. 
			At this stage of the estimation process, the parameters are assumed to be constant with the wavenumbers.
			\item In the \textbf{\glsreset{gn}\gls{gn}} step, we process the raw acquisition $\mathbf{y}\in\mathbb{R}^{N_a}$ to obtain the final estimation $\hat{\bm{\beta}}$ of the complete set of parameters.
			To achieve this, we initialize a \gls{lm} algorithm~\cite{More78} with the parameter vector $\bm{\beta}^{[0]}$ whose elements were inferred in the previous steps.
			Subsequently, we iterate through the algorithm to solve eq.~\eqref{eq:cost}.
		\end{itemize}
	
		The following sections describe each of these steps in further detail.
		It is important to note that in certain scenarios, such as non-imaging systems where a one-dimensional acquisition is obtained using a single pixel sensor, the mean over neighbors and flat field statistic may not be available.
		Nonetheless, in these cases, the algorithm can still be applied by setting $\mathbf{u}$ equal to $\mathbf{y}$ and the elements of $\mathbf{w}$ equal to the average value of $\mathbf{y}$.
	
	\subsection{Step 1: Gain estimation}
	\label{ssec:characterization_gain}
	
		We formalize the problem associated to the gain estimation as follows:
		\begin{equation}
			\hat{\mathbf{a}}=\arg\min_{\mathbf{a}}\sum_{i=1}^{N_a}(\mathcal{A}(\sigma_i)-w_i)^2\,,
			\label{eq:cost_gain}
		\end{equation}
		where $\hat{\mathbf{a}}=\{\hat{a}_m\}_{\range{m}{0}{N_d}}$ describes our estimation the coefficients of the polynomial representation  $\hat{\mathcal{A}}(\sigma)=\sum_{m=0}^{N_d}\hat{a}_m\sigma^m$ of the gain $\mathcal{A}(\sigma)$. The vector $\mathbf{w}=\{w_i\}_{\range{i}{1}{N_a}}$ contains the samples of the flat field statistic.
		
		We propose to solve the problem above with a nonlinear regression approach using the \gls{lm} algorithm with the implementation of~\cite{More78}, as described in Appendix~\ref{ssec:lm}. We initialize the gain coefficients vector $\mathbf{a}$ by setting the first element to the average value of $\mathbf{w}$ and the rest to zero.
		
	\subsection{Step 2: \texorpdfstring{\Glsentrylong{ml}}{Maximum likelihood} initialization}
	\label{ssec:characterization_ml}
	
		The \glsentryfull{ml} initialization step defines a procedure that is as an extension of the simplistic \gls{opd} estimation algorithm proposed in~\cite{Dole19}. 
		In the step, we assume to operate in a low finesse regime, so that the transmittance response $T_{\bm{\beta}}(\sigma)$ behaves like the 2 waves model of eq.~\eqref{eq:airy_1}: 
		\begin{equation}
			T_{\bm{\beta}}(\sigma)=\left(1+\frac{2r_0^{}}{1+r_0^2}\cos(2\pi\delta\sigma-\varphi^{}_0)\right)\mathcal{A}(\sigma)\,.
		\end{equation}
		In the above equation, we implicitly impose that the reflectivity $\mathcal{R}$ is uniform and equal to $r_0$ over the whole wavenumber range.
		By normalizing both terms of the minimization problem of eq.~\eqref{eq:cost} and applying it to the mean over neighborhood vector $\mathbf{u}$, the problem can be rewritten as:
		
			\begin{align}
				\hat{\bm{\beta}}&\approx\arg\min_{\bm{\beta}}\sum_{i=1}^{N_a}\left(\frac{T_{\bm{\beta}}(\sigma_i)}{\mathcal{A}(\sigma_i)}-\frac{u_i}{\hat{\mathcal{A}}(\sigma_i)}\right)^2
				\approx\arg\min_{\bm{\beta}}\sum_{i=1}^{N_a}\left(\alpha\cos\left(2\pi\delta\sigma_i-\varphi_0^{}\right)-v_i\right)^2\,,\label{eq:cost_ml_2}
			\end{align}
		where we defined $\alpha:=2r_0^{}/(1+r_0^2)$ and $v_i:=\left(u_i-\hat{\mathcal{A}}(\sigma_i)\right)/\hat{\mathcal{A}}(\sigma_i)$, assuming $\mathcal{A}(\sigma_i)\approx\hat{\mathcal{A}}(\sigma)$.
		
		Eq.~\eqref{eq:cost_ml_2} is in the form of the classical problem of the inference of the parameters in a sinusoid affected by Gaussian noise, which is a well known problem in the literature~\cite[Example 7.16]{Kay93}.
		Specifically, it is a well known result that the maximum likelihood estimator $\hat{\delta}$ is equal to the \gls{opd} value which maximizes the \textit{periodogram}:
		\begin{equation}
			\hat{\delta}=\arg\max_{\delta\in\left[0,\frac{1}{2\Delta\sigma}\right]}\left\lvert\sum_{i=1}^{N_a}v_i\exp(-j2\pi\delta\sigma_i)\right\rvert\,.
			\label{eq:delta_zero}
		\end{equation}
		In other terms, the estimator is the value that maximizes the generalized \gls{dft} of $\mathbf{v}$.
		The above result is valid for values of $\delta$ reasonably far from the extremes of the interval $[0, 1/(2\Delta\sigma)]$, where $\Delta\sigma=(\sigma_{max}-\sigma_{min})/N_a$ is the average central wavenumber step.
		
		Eq.~\eqref{eq:delta_zero} can be solved numerically over a sampled version of the interval of interest, yet the accuracy is limited to a resolution of $1/(2N_a\Delta\sigma)$.
		If the \gls{opd} is approximately known, it is computationally efficient to evaluate eq.~\eqref{eq:delta_zero} within a reduced interval centered around its nominal value. 
		
		The estimation $\hat{r}_0$ of the reflectivity $r_0$ is then obtained in terms of the the estimation $\hat{\alpha}$ of the amplitude $\alpha$ of the sinusoid:
		\begin{equation}
			\hat{r}_0=1-\sqrt{1-\hat{\alpha}^2}\,,\;\;\;\;\;\;\textrm{where }\;\;\;\hat{\alpha}=\frac{2}{N_a}\left\lvert\sum_{i=1}^{N_a}v_i \exp(-j2\pi\hat{\delta}\sigma_i)\right\rvert\,,\label{eq:reflectivity_hat}
		\end{equation}		
		 and the estimation $\hat{\varphi}^{}_0$ of $\varphi^{}_0$ is:
		\begin{equation}
			\hat{\varphi}_0^{}=\arctan\frac{\sum_{i=1}^{N_a}v_i\sin(2\pi\hat{\delta}\sigma_i)}{\sum_{i=1}^{N_a}v_i\cos(2\pi\hat{\delta}\sigma_i)}\,.
			\label{eq:phase_shift_hat}
		\end{equation}
		In the above equation, $\arctan$ denotes the four-quadrant arctangent version that allows for $\hat{\varphi}_0^{}$ to assume any value in the range $[-\pi,\pi)$. The \gls{ml} method requires very low computational power, but its applicability is limited by the validity of its assumptions. Some other possible initialization strategies, such as the \gls{es} developed in~\cite{Pico20} which is based on a grid search in the sample space of the parameters, have the advantage to work with a wider variety of models. They are however vastly slower and may not necessarily produce more accurate results, as the estimations for $r_0^{}$ and $\varphi_0^{}$ are limited to the finite amount of values of the discrete sample space. 
	
	\subsection{Step 3: Trust region refinement (TRR)}
	\label{ssec:characterization_gna}
	
		The final parameter estimation $\hat{\bm{\beta}}$ follows a similar procedure as described in Section~\ref{ssec:characterization_gain}. 
		Specifically, the \gls{lm} algorithm is employed once again, but this time to solve eq.~\eqref{eq:cost}.
		The parameter vector $\bm{\beta}$ is initialized with values from $\bm{\beta}^{[0]}=[\hat{a}_0,...,\,\hat{a}_{N_d},\, \hat{r}_0^{},\, 0, ...,\,0,\, \hat{\delta},\hat{\varphi}_0^{}]$, where the non-zero elements correspond to the coefficients estimated at the step 1 and 2 of the algorithm.

\section{Experimental results}
\label{sec:experiments}

	This section presents the experimental results obtained from the characterization of a series of \gls{imspoc} prototypes with various characteristics. In Section~\ref{ssec:experiments_setup} we describe the experimental setup, in Section~\ref{ssec:experiments_robustness} we test various configurations for the proposed algorithm and compare its performances with previous works. Finally, in Section~\ref{ssec:experiments_discussion} we discuss the physical interpretation of the parameters. A Python implementation of the proposed algorithms, together with a simulator of the image formation for multi-aperture Fabry-Perot imaging spectrometers, is available at the first author's repository~\cite{web_IRCA}.

	\subsection{Experimental setup}
	\label{ssec:experiments_setup}
	
		\begin{table}[t]
		    \centering
		    \sisetup{detect-weight=true,detect-inline-weight=math}
		    \footnotesize
		    \NineColors{saturation=high}
		    \begin{talltblr}[
		    	caption = {Characteristics of the available \gls{imspoc} prototypes used in this work and of their spectral characterization experimental acquisitions.
		    	\label{tab:proto}},
		    	note{*} = {In this prototype, two interferometers are both at the optical contact for testing purposes.},
		    	note{**} = {This cell gives the mean and standard deviation of the step size, as the wavenumber space is irregularly spaced for this experiment.}]
		    	{
		    		colspec = {l|ccccccc},
		    		vlines = {white},
		    		hlines = {white},
		    		cell{2}{1} = {red3, fg=yellow9, font=\bfseries},
		    		cell{1-2}{2-8} = {red3, fg=yellow9, font=\bfseries},
		    		cell{4,6}{2-8} = {red9},
		    		cell{3-6}{1}={red3, fg=yellow9},
		    		cell{3-6}{1} = {font=\bfseries},
		    	}
		    	&\SetCell[c=4]{c}{Device specifications}&\SetCell[c=3]{c}{Acquisition specifications}\\
		    	Prototype label& \makecell{Interf.s\\$N_i$} & \makecell[c] {$\Delta d$\\$[\si{nm}]$} & \makecell[c]{Focal plane\\size [\si{px}]} & \makecell[c]{Subimage\\size [\si{px}]} & \makecell[c]{Wavenumber\\range [\si{mm^{-1}}]} & \makecell{Acq.s\\$N_a$} & \makecell[c]{$\Delta\sigma$\\$[\si{{cm}^{-1}}]$}
		    	\\
		    	\Glsentryshort{p1} &$216$ &$100$ &$1096\times2808$ &$100\times 100$ &$1000-2000$ &101  &100
		    	\\
		    	\Glsentryshort{p2} &$319$ &$87.5$ &$1096\times2808$ &$96\times 96$   &$1000-2850$ &721 &25
		    	\\
		    	\Glsentryshort{p5} &$672$ &$87.5$ &$1096\times2808$ &$66\times66$  &$1230-2880$ &551 &30
		    	\\
		    	\Glsentryshort{p3}  &$79\,(+1)$\TblrNote{*} &$200$ &$512\times640$   &$64\times 64$   &$625-1000$  &343 &$11 \pm 12$\TblrNote{**}
		 	\end{talltblr}
		\end{table}
	
		For this work, the characterization datacubes were captured with the setup shown in \figurename~\ref{fig:experimental_setup}, using a tunable monochromatic light source from Zolix Instruments Co., Ltd, with a 500 \si{W} Xenon light source model Gloria-X500A and a monochromator model Omni-300$\lambda$i. We also utilized a 5.3-inch diameter integrating sphere coated with Spectralon (model 4P-GPS-053-SF from Labsphere, Inc.). The incident optical power was measured either with the fiber optic gated spectrometers model USB2000+ from Ocean Optics, Inc. or with the photodiode power sensor model S120VC from Thorlabs. The product specifications can be found on the websites of the respective manufacturers.
		
		The devices under test are four different \gls{imspoc} prototypes, whose characteristics are described in \tablename~\ref{tab:proto}.
		Each prototype features an array of interferometers disposed over a bidimensional matrix in a staircase pattern, whose thicknesses linearly increase with a nominally constant step size $\Delta d$.
		While sharing the same underlying concept, each prototype is specifically designed for different applications. \Glsentryshort{p1} and 2 are specifically tailored for the measure of atmospheric pollution, \glsentryshort{p5} functions as an imaging system for capturing the phenomenon of northern lights, and \glsentryshort{p3} is intended for greenhouse gas detection. In terms of spectral sensitivity, prototypes 1 to 3 operate within the visible/ultraviolet wavelength range, whereas \glsentryshort{p3} covers the near-infrared spectrum.
		For each device, a characterization datacube was captured using the procedure described in Section~\ref{ssec:characterization_setup}. The central wavenumbers of the monochromator are chosen to be regularly spaced with a step size $\Delta\sigma$. 
		The wavenumber step size $\Delta\sigma$ is selected to satisfy the Nyquist condition $\Delta\sigma<1/(2\delta_{max})$. This selection ensures that aliasing effects are avoided in the sampling of the transmittance response of the interferometer with the largest \glsentryshort{opd} $\delta_{max}=2(N_i-1)\Delta d$. The condition is satisfied in all experimental setups, although only by a small margin for \glsentryshort{p1} and to a lesser extent for \glsentryshort{p5}, due to time constraints.
		The specifications for these measurements are reported in~\tablename~\ref{tab:proto}.
		
		\begin{table}[t]
		    \centering
		    \caption[Model characterization results]{Model characterization \gls{rmse} comparison. Best results are in bold.}
		    \footnotesize
		    \NineColors{saturation=high}
		
			\begin{tblr}
			{
				colspec = {lcccccc},
				vline{2-7} = {white},
				hline{2,3,6,8,10} = {white},
				cell{1}{2-7} = {red3, fg=yellow9},
				cell{3,5,7,9}{3-7} = {red9},
				cell{2-10}{2} = {red9},
				cell{2-10}{1}={red3, fg=yellow9},
				cell{2-10}{2} = {font=\bfseries},
			}
				&\textbf{Method}&$W$&\textbf{\Glsentryshort{p1}}&\textbf{\Glsentryshort{p2}}&\textbf{\Glsentryshort{p5}}&\textbf{\Glsentryshort{p3}}\\
				\SetCell[r=6]{c}{\rotatebox[origin=c]{90}{\textbf{Fixed} $\mathcal{A}$}}
				&\glsentryshort{ml}~\cite{Dole19}&$2$& $0.3022 \pm 0.0605$ &$0.1948 \pm 0.0720$ & $0.4363 \pm 0.2338$ & $0.2291 \pm 0.0756$ \\
				&\SetCell[r=3]{c}{\glsentryshort{es}~\cite{Pico20}}& $2$& $0.3052 \pm 0.0638$ &$0.2319 \pm 0.0792$ & $0.4654 \pm 0.2393$ & $0.2472 \pm 0.0795$ \\
				&& $3$ & $0.2941 \pm 0.0638$ &$0.2153 \pm 0.0991$ & $0.4672 \pm 0.2403$ & $0.2472 \pm 0.0800$ \\
				&& $\infty$ & $0.2934 \pm 0.0640$ &$0.2204 \pm 0.1247$ & $0.4673 \pm 0.2404$ & $0.2472 \pm 0.0800$ \\
				&\SetCell[r=2]{c}{\glsentryshort{ml}+\glsentryshort{gn}}& $2$& $0.2721 \pm 0.0637$ &$0.2698 \pm 0.2166$ & $0.4153 \pm 0.2338$ & $0.1856 \pm 0.0852$ \\
				&& $\infty$ & $0.2544 \pm 0.0631$ &$0.2445 \pm 02217$ & $0.4126 \pm 0.2342$ & $0.1836 \pm 0.0873$ \\
				\SetCell[r=3]{c}{\rotatebox[origin=c]{90}{\textbf{Free} $\mathcal{A}$}} 
				&\SetCell[r=2]{c}{\glsentryshort{ml}+\glsentryshort{gn}}& $2$ & $0.2169 \pm 0.0392$ &$0.1691 \pm 0.0366$ & $0.2186 \pm 0.0528$ & $0.0724 \pm 0.0210$ \\
				&& $\infty$ & $\mathbf{0.1937 \pm 0.0432}$ &$0.1336 \pm 0.0343$ & $0.2170 \pm 0.0519$ & $\mathbf{0.0669 \pm 0.0222}$ \\
				&\glsentryshort{es}+\glsentryshort{gn}& $\infty$& $\mathbf{0.1937 \pm 0.0432}$ &$\mathbf{0.1335 \pm 0.0343}$ & $\mathbf{0.2130 \pm 0.0520}$ & $0.0676 \pm 0.0233$ \\
			\end{tblr}
		
		    \label{tab:fitting}
		    \end{table}

		Given a characterization datacube and a specific subimage within it, the central pixel of the chosen subimage is extracted to construct the raw acquisition vector $\mathbf{y}$. The mean over neighbors is computed using a $11 \times 11$ kernel window centered around the extracted pixel, while the 90-percentile metric is instead employed as the flat field statistic. 
		Next, we apply the characterization method described in Section~\ref{sec:characterization} to obtain the characterization vector $\hat{\bm{\beta}}$, with $N_d=5$ as the degree of the polynomial for the reflectivity and gain. 
		To verify the quality of the estimation, we use the \gls{rmse} metric, defined as follows:
		\begin{equation}
			RMSE=\left[\frac{1}{N_a}\sum_{i=1}^{N_a}\left(\frac{T_{\hat{\bm{\beta}}}(\sigma_i)-y_i}{\overline{y}}\right)^2\right]^{1/2}\,.
			\label{eq:rmse}
		\end{equation}
		where $\overline{y}=(\sum_{i=1}^{N_a}y_i)/N_a$ denotes the mean value of $\mathbf{y}$ and $T_{\hat{\bm{\beta}}}(\sigma_i)$ is eq.~\eqref{eq:transfer_function} evaluated with the estimated vector of parameters $\hat{\bm{\beta}}$. This metric serves as benchmark for comparing with the other characterization methods being tested.
		We then repeat this procedure in order to characterize the transmittance response of the central pixels for all $N_i$ interferometers of the device.
	
	\subsection{Algorithm and model comparisons}
	\label{ssec:experiments_robustness}
	
		The \gls{irca} is tested here with different configurations, and we assume that the gain estimation is always carried as a pre-processing step. We compare its results with previous works~\cite{Dole19, Pico20} which we can conveniently frame within our proposed framework.
		
		\begin{figure*}
			\captionsetup[subfigure]{justification=centering}
			\centering
			\begin{subfigure}[b]{0.49\linewidth}
				\centering
				\includegraphics[width=\linewidth]{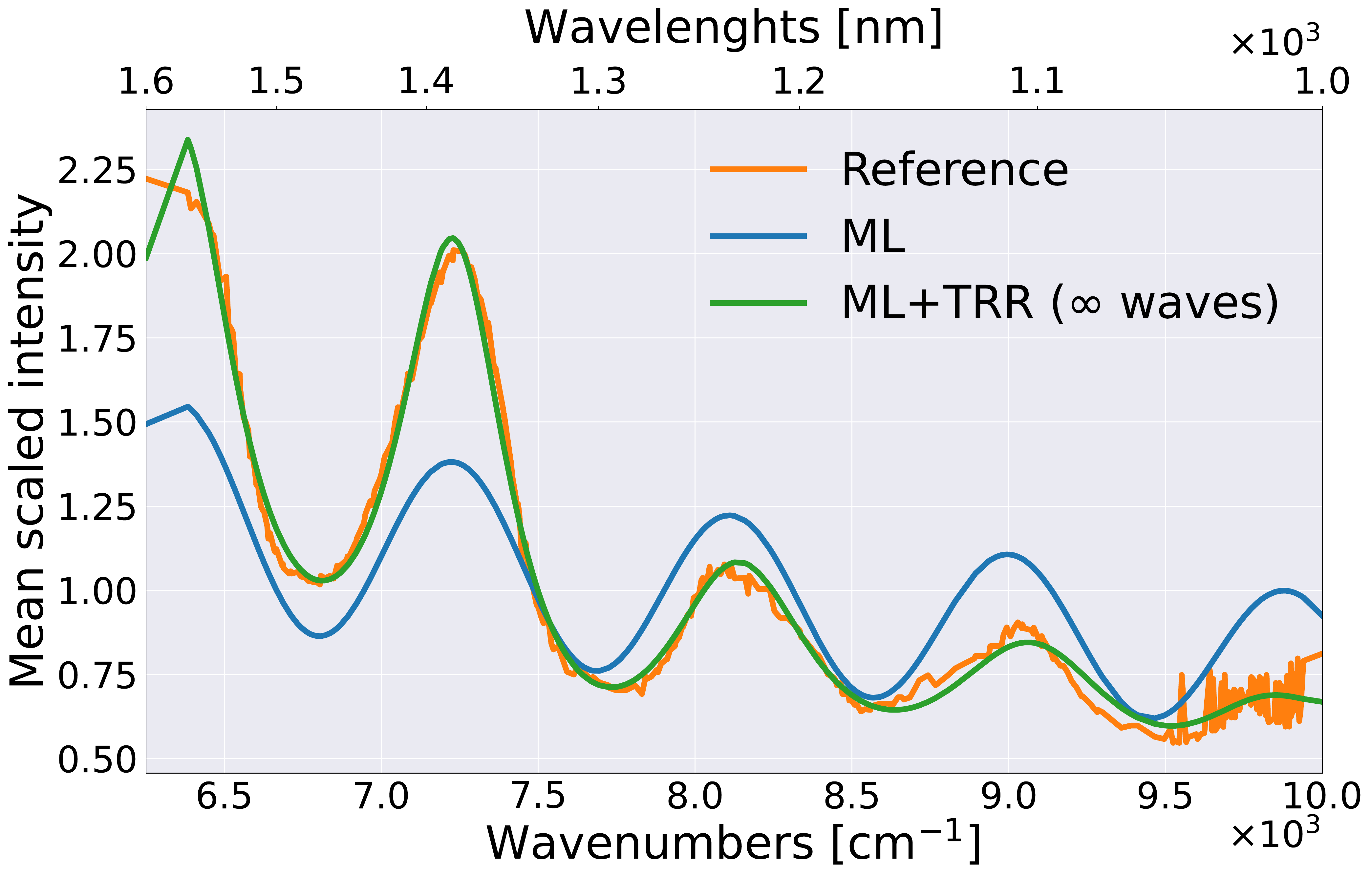}
				\caption{\Glsentryshort{p3}: transmittance response of interferometer \#20.}
				\label{fig:curves_p3}
			\end{subfigure}
			\hfil
			\begin{subfigure}[b]{0.49\linewidth}
				\centering
				\includegraphics[width=\linewidth]{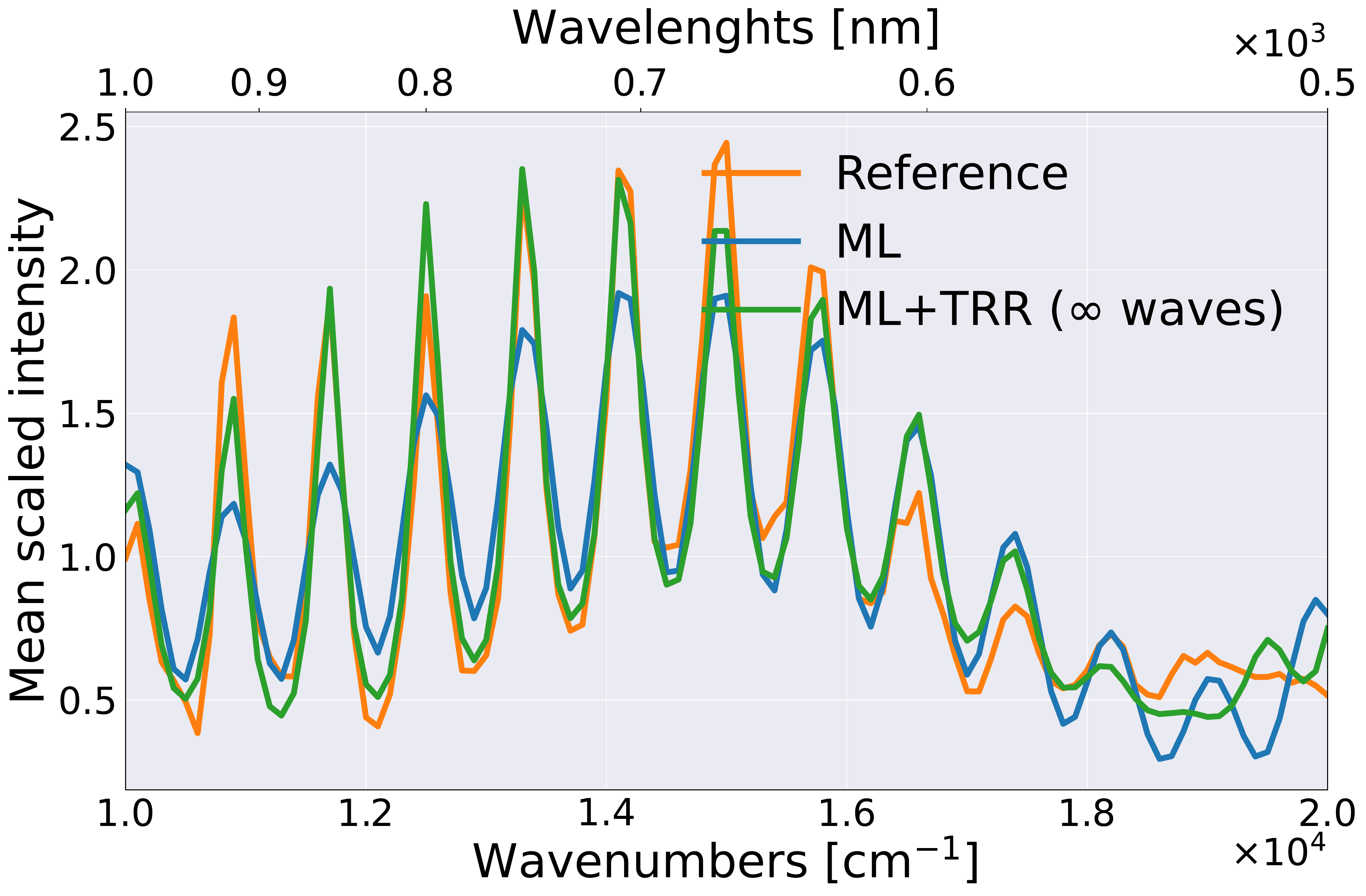}
				\caption{\Glsentryshort{p1}: transmittance response of interferometer \#50.}
				\label{fig:curves_p1}
			\end{subfigure}
			
			\smallskip
				
			\begin{subfigure}[b]{0.49\linewidth}
				\centering
				\includegraphics[width=\linewidth]{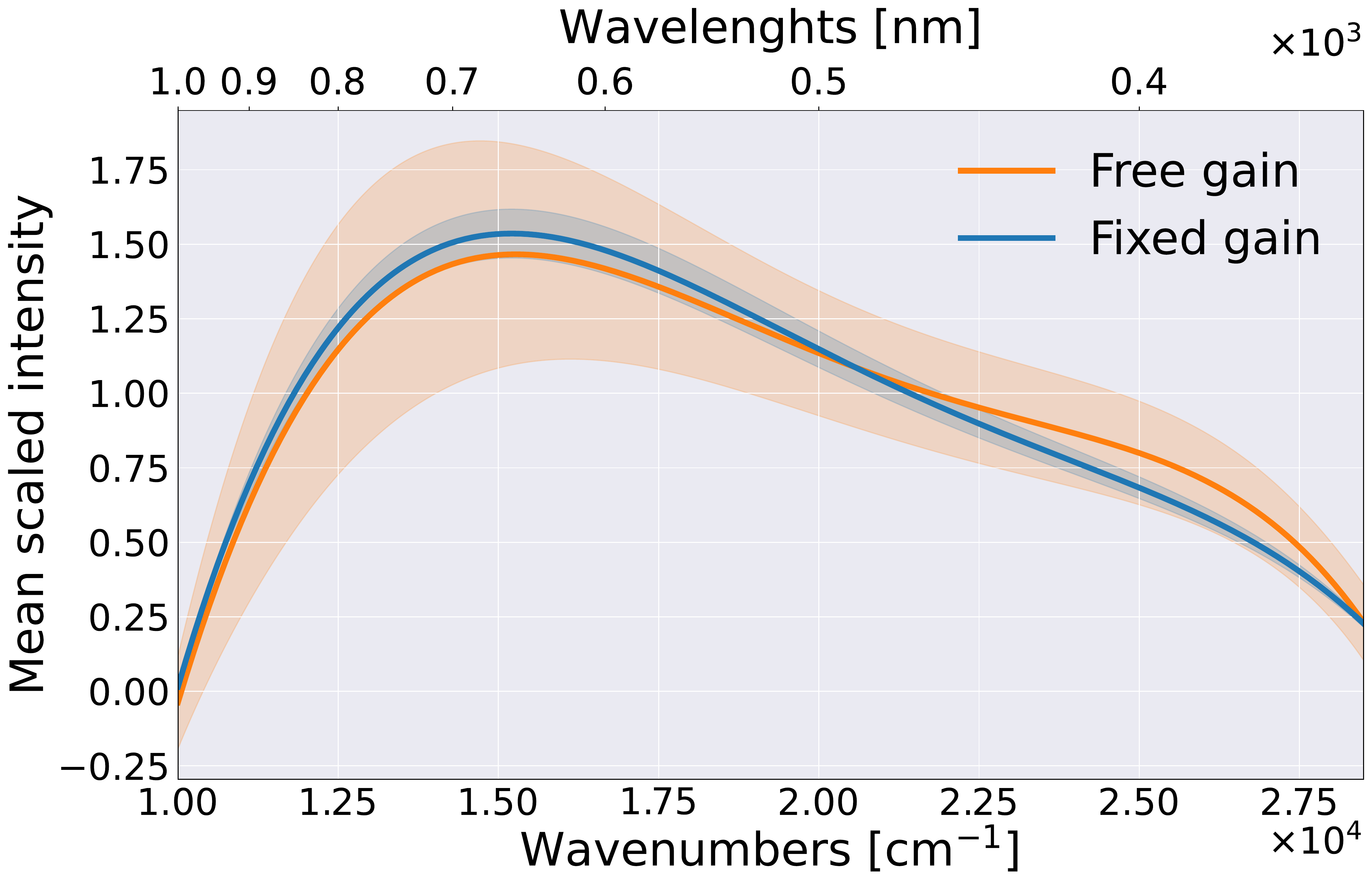}
				\caption{\Glsentryshort{p2}: gain estimation comparison.}
				\label{fig:curves_gain}
			\end{subfigure}
			\hfil
			\begin{subfigure}[b]{0.49\linewidth}
				\centering
				\includegraphics[width=\linewidth]{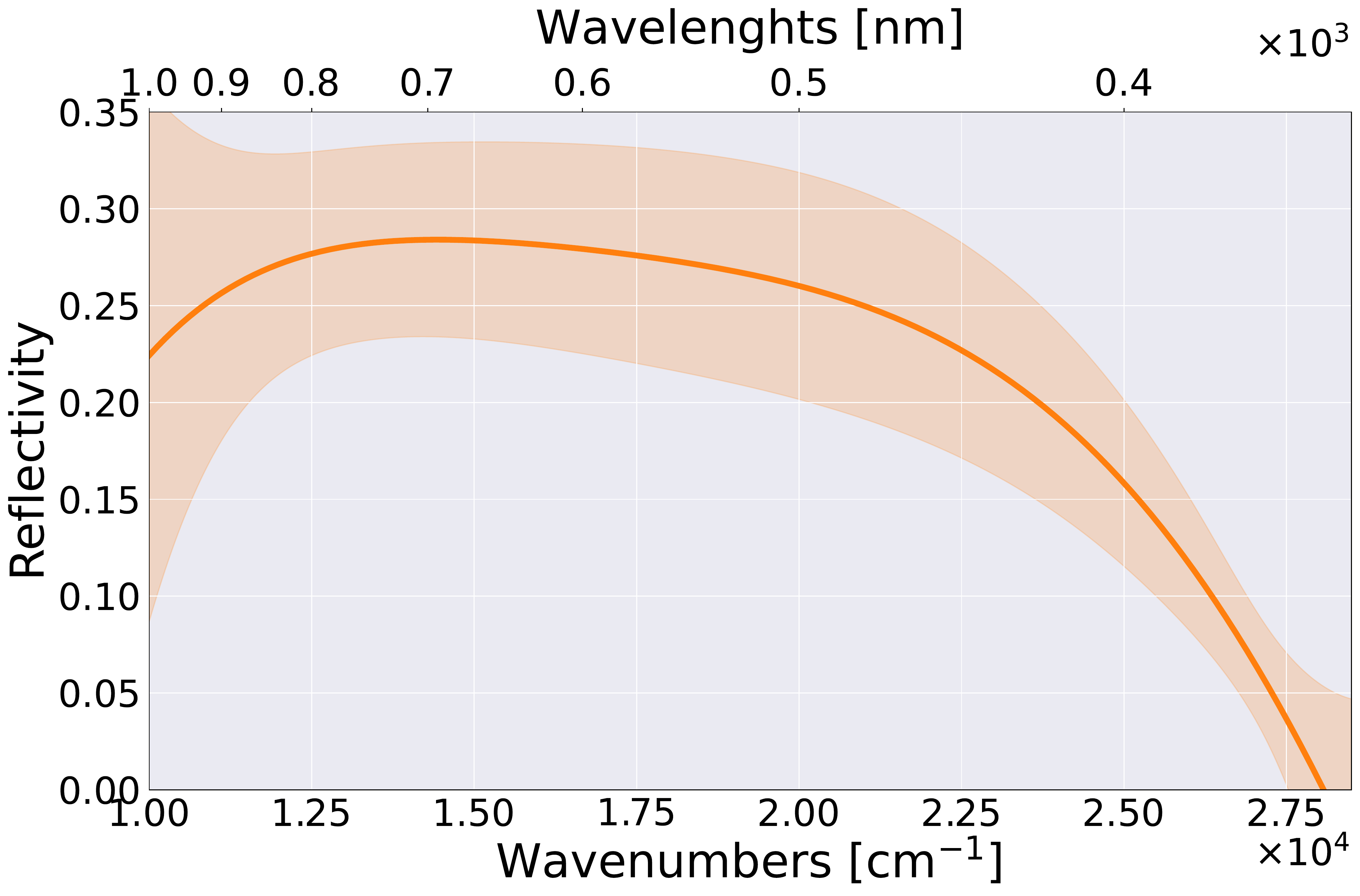}
				\caption{\Glsentryshort{p2}: estimated reflectivity.}
				\label{fig:curves_reflectivity}
			\end{subfigure}
			
			\caption[Spectral response fitting]{
				Spectral responses of the characterization results with the \gls{irca}. (\subref{fig:curves_p3}), (\subref{fig:curves_p1}) Two examples of fitting the acquisition with the transmittance response using estimated parameters. The acquisition (in orange) is compared to the estimation with the proposed \gls{irca} method (in green) and with the \glsentryshort{ml} method (in blue). (\subref{fig:curves_gain}) Estimated gain. In the orange curve, we allow the \gls{lm} algorithm to freely update the parameters of the gain polynomial, while in the blue curve we only allow to change the scaling factor of the initialized gain. (\subref{fig:curves_reflectivity}) Estimated reflectivity. The shaded area around the curves represents one standard deviation interval across different interferometers.
			}
			\label{fig:curves}
		\end{figure*}
		
		We employ different wave models for the optical transmittance response $T_{\bm{\beta}}(\sigma)$, according to the definitions of eq.s~\eqref{eq:airy} and~\eqref{eq:transfer_function}, specifically for the case of $W=2$, $3$, or $\infty$ emerging light rays.

		The \gls{rmse} results, given in \tablename~\ref{tab:fitting}, shows that, when all the three steps are performed, the proposed method is consistently the best performing, regardless of the different characteristics of the prototypes.
		It also highlights how the $\infty$-wave model, which is a better representation of the generalized Airy distribution, provides a more accurate fit for the spectral response.
		Both tested initializations reach comparable results, suggesting that the \gls{ml} method should be preferred, as it is faster by a factor of 10-20 times over the \gls{es} methodology.
		
		The proposed procedure was tested both with and without the trust region refinement step, in order to showcase the advantage of the iterative curve fitting procedure.
		The \gls{gn} step has a considerable impact on the accuracy of the results. This is due to the \gls{lm} algorithm exploring a continuous space of parameters, resulting in better performance with respect to the \gls{ml} by itself where the \gls{opd} space is explored in discrete steps.
		A visual comparison between their reconstructed spectral responses is shown in \figurename~\ref{fig:curves}. While the \gls{ml} algorithm by itself already infers the \gls{opd} with a remarkable accuracy, the curve does not follow accurately the trend of the data, as the other parameters vary significantly with the wavenumbers.
		
		The results for the \glsentryshort{p2} for the estimation of the gain and reflectivity are given in \figurename~\ref{fig:curves}. The figures showcase the advantage of the \gls{gn} step in assessing the gain for each detector separately.
		The analysis of the reflectivity reveals a reduced sensitivity of the instrument at the extreme values of the wavenumber range, which aligns with the expected spectral response of the reflective coating.
		
		\begin{figure*}
		    \centering
		    \begin{subfigure}[b]{0.7\linewidth}
		    	\includegraphics[trim={0 0 0 0}, clip, width=0.99\linewidth]{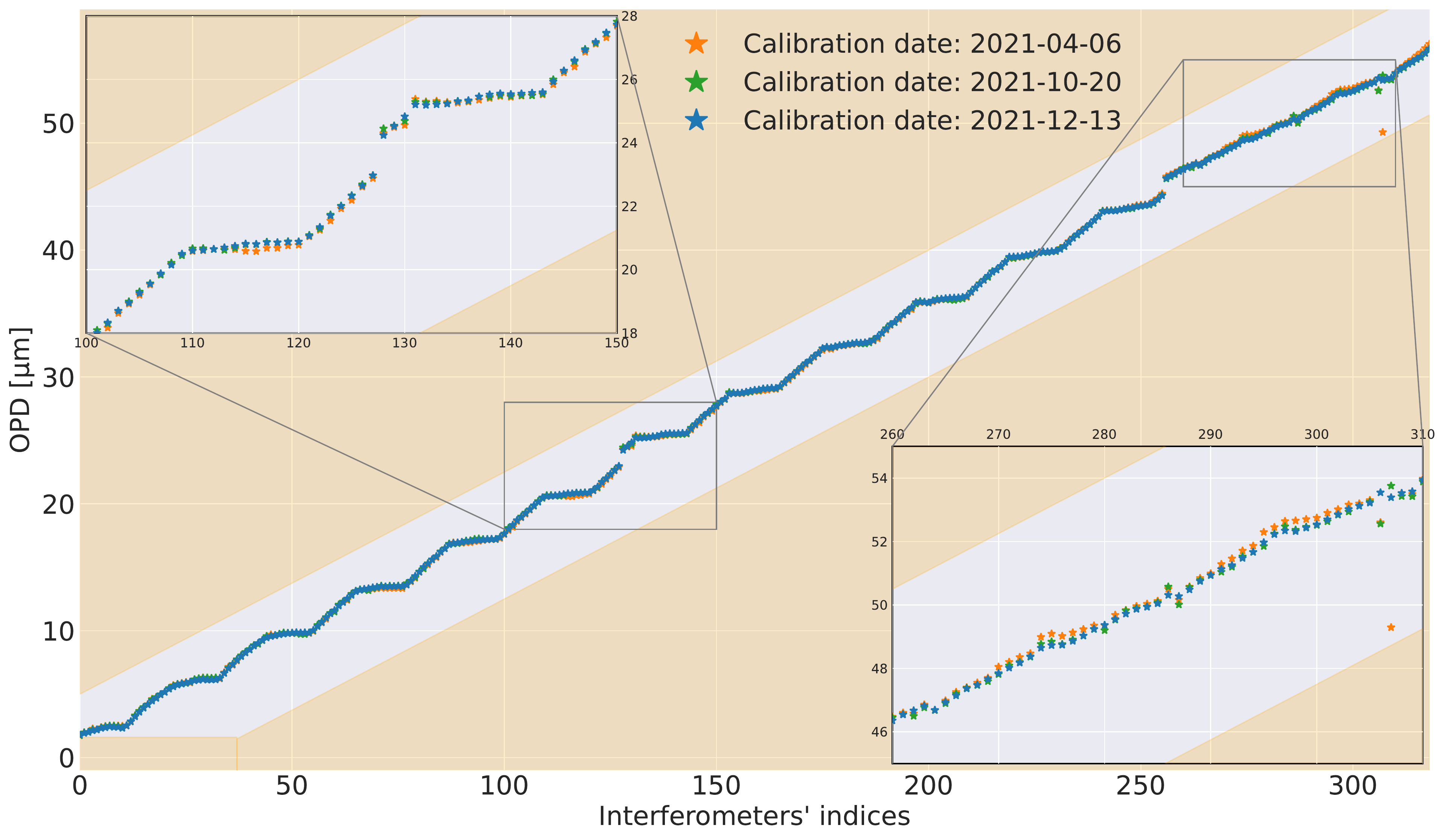}
		    	\caption{Comparison of estimated \glspl{opd} (with zoomed regions).}
		    	\label{fig:opd_comparison}
		    \end{subfigure}
		    \hfil
		    \begin{minipage}[b]{.29\linewidth}
		    	\begin{subfigure}[b]{\linewidth}
		    		\centering
		    		\includegraphics[width=\linewidth]{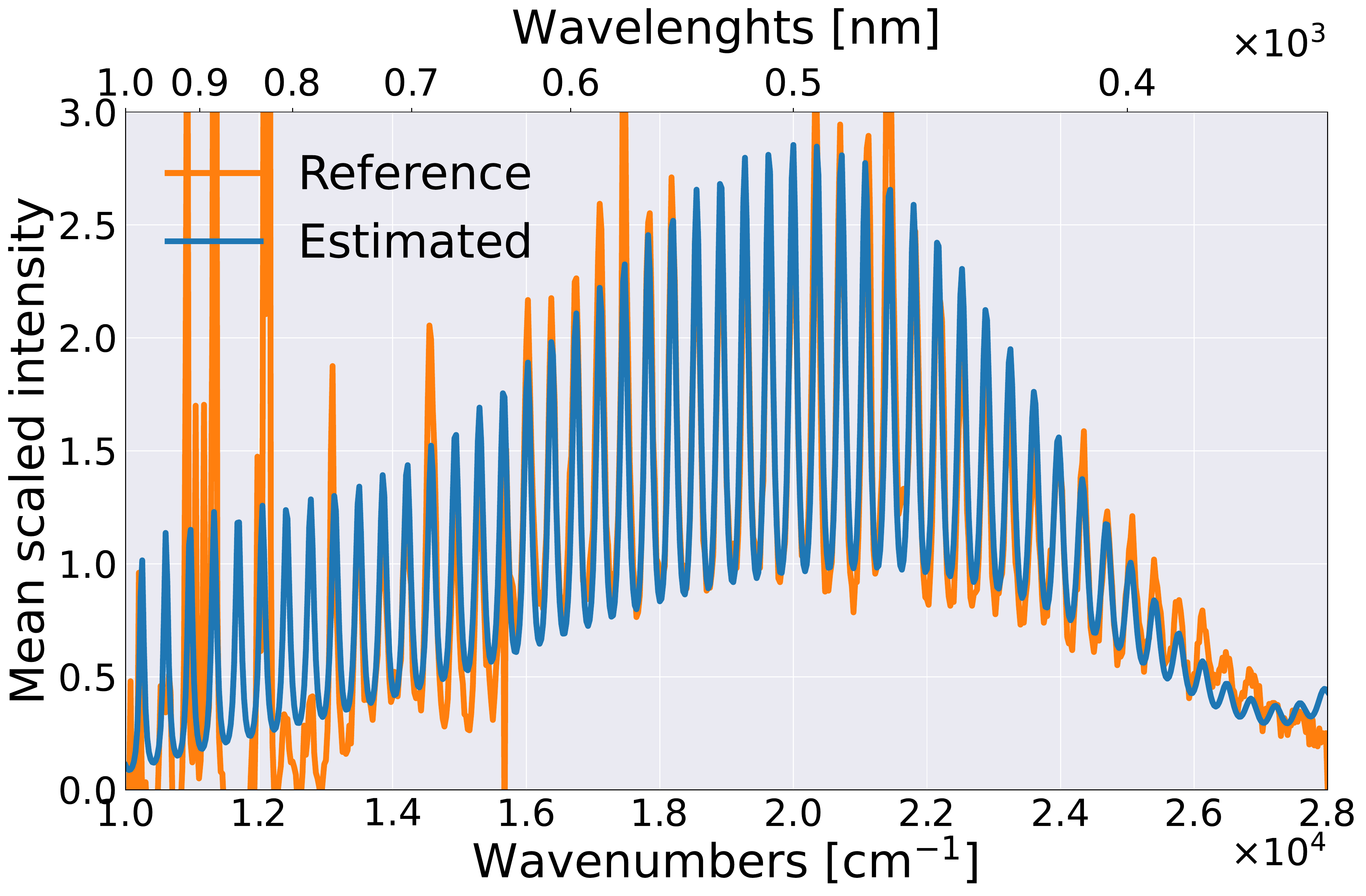}
		    		\caption{Calibration date: 2021-10-20.}
		    		\label{fig:opd_old}
		    	\end{subfigure}
		    	
		    	\smallskip
		    	
		    	\begin{subfigure}[b]{\linewidth}
		    		\centering
		    		\includegraphics[width=\linewidth]{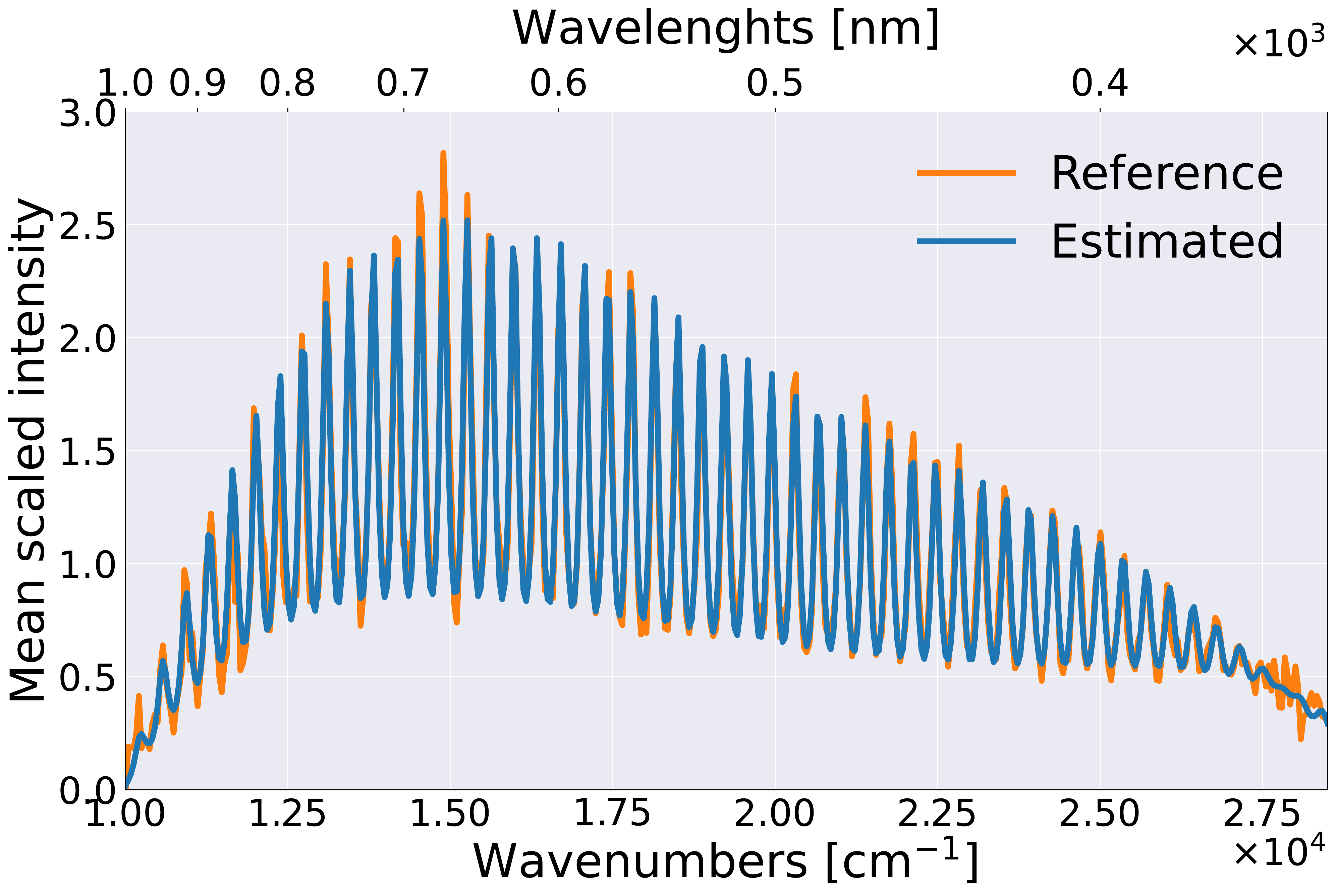}
		    		\caption{Calibration date: 2021-12-13.}
					\label{fig:opd_new}
		    	\end{subfigure}
		    \end{minipage}
			
		    \caption[\Glsentryshort{opd} estimation]{
		    	Characterization of \glsentryshort{p2} from measurements taken at different times. (\subref{fig:opd_comparison}) Visualization of the estimated \glspl{opd} for all the interferometers, arranged in ascending order of nominal thicknesses. The yellow halo indicates the region that was not explored for the \gls{ml} initialization, as that strays too far from the nominal values known from the design of the instrument.
		    	(\subref{fig:opd_old}), (\subref{fig:opd_new}) Comparison between the estimated transmittance responses (blue) and raw acquired data (orange) for interferometer \#150. 
				There are significant differences in the intensity values and appearance of the orange curves between the two figures. This stems from the fact that the raw acquisitions on 2021-12-23 were divided by the spectral response of the light source, while the acquisitions on 2021-10-20 are unaltered as we lack this information for that particular session.
			}
		    \label{fig:opd}
		\end{figure*}
		
	\subsection{Physical interpretation of the results}
	\label{ssec:experiments_discussion}
		
		For the \glsentryshort{p2}, the measurements for the characterization was repeated at three different dates, using progressively refined setups. 
		The proposed method was applied to each of those datasets in order both to analyze the robustness of the algorithm and to detect eventual drifting in the parameters.
		\figurename~\ref{fig:opd} provides a visual comparison of the results.
 		For the normal incidence of the light illumination, the \glspl{opd} are roughly expected to be double the thickness of interferometers, as a consequence of substituting $\theta=0$ in eq.~\eqref{eq:opd_fabry_perot}.
		The estimation of the \glspl{opd} in \figurename~\ref{fig:opd_comparison} stays vastly consistent across all sessions, with only very sparse examples where the results do not align.

		\begin{figure*}
			\captionsetup[subfigure]{justification=centering}
			\centering
			\begin{subfigure}[b]{0.70\linewidth}
				\includegraphics[trim={0 0 0 0}, clip, width=0.99\linewidth]{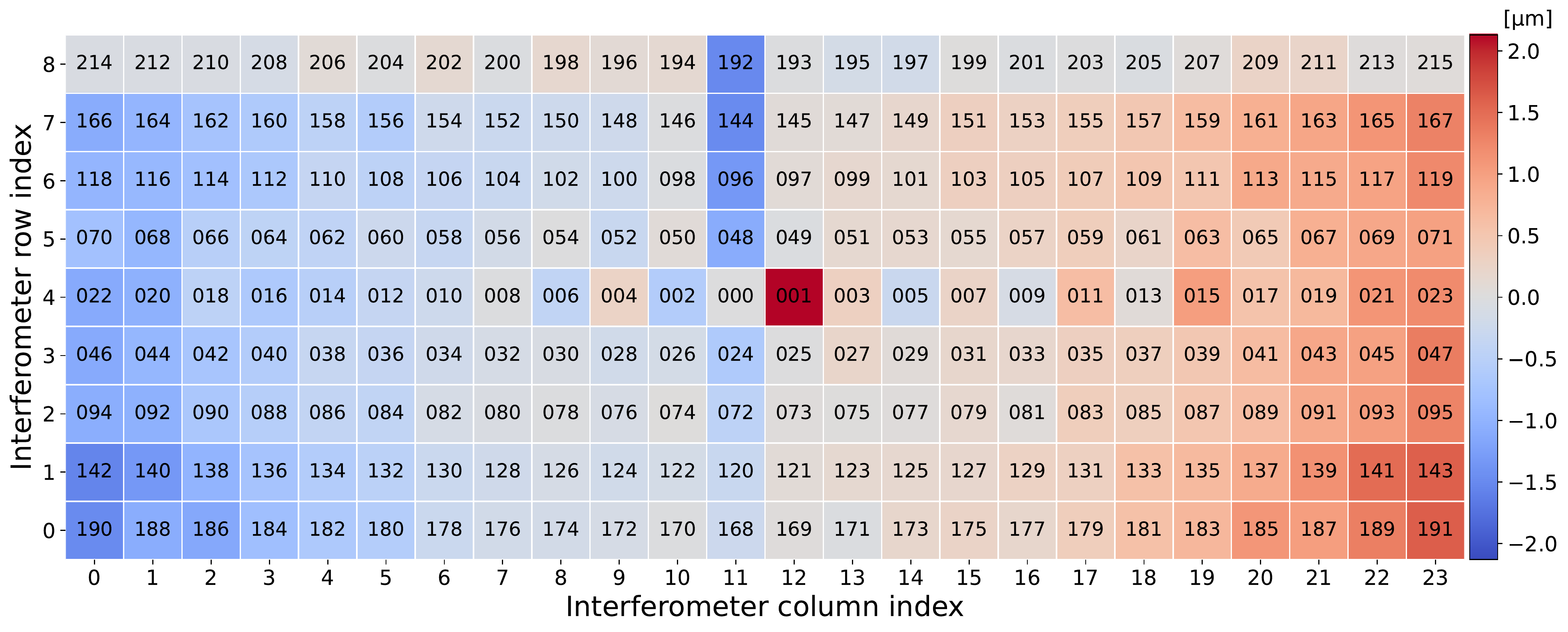}
				\caption{\Gls{opd} step difference deviation from nominal value.}
				\label{fig:plate_data}
			\end{subfigure}
			\begin{minipage}[b]{.29\linewidth}
				\begin{subfigure}[b]{\linewidth}
					\centering
					\includegraphics[width=\linewidth]{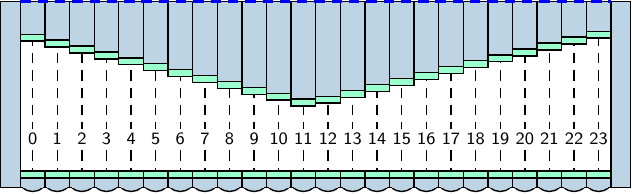}
					\caption{Ideal optical plate installation.}
					\label{fig:plate_ideal}
				\end{subfigure}
				
				\vspace{5mm}
				
				\begin{subfigure}[b]{\linewidth}
					\centering
					\includegraphics[width=\linewidth]{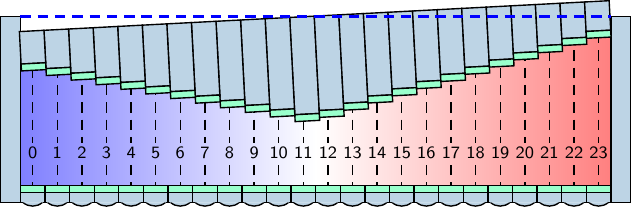}
					\caption{Tilted plate.}
					\label{fig:plate_tilted}
				\end{subfigure}
			\end{minipage}
			\caption[\Glsentryshort{opd} plates]{
				Estimated \glsentryshortpl{opd} for the \glsentryshort{p1}.
				(\subref{fig:plate_data}) The heatmap presents the variation in the increase of the \gls{opd} between successive interferometers. The indexes of the interferometers are arranged in ascending order of their nominal thickness. (\subref{fig:plate_ideal}), (\subref{fig:plate_tilted}) Effect of tilting the optical plate on with respect to an aligned one. The tilt can result in either a compression (indicated by red shades) or an expansion (indicated by blue shades) of the air gap in the cavity with respect to the nominal value.  The indices inside the cavities denote the positions of the interferometers within the row.
			}
			\label{fig:plate}
		\end{figure*}
		The figure shows a pattern of alternating slopes. To make sense of this effect, we also present the estimation of the \glspl{opd} for the \glsentryshort{p1}.
		\Glsentryshort{p1} is designed with a distinctive staircase pattern, characterized by cavity thicknesses that increase in constant steps on both sides of a central vertical axis. 
		This pattern is reflected in the deviation of the estimated \gls{opd} increase with respect to the expected value. As depicted in \figurename~\ref{fig:plate}, the observed discrepancy can likely be attributed to a tilt of the optical plates relative to the desired parallel installation.
		
		%
		
\begin{figure*}
			\centering
			
			\hspace*{0.01\linewidth}
			\begin{subfigure}[b]{0.91\linewidth}
				\centering
				\includegraphics[width=\linewidth]{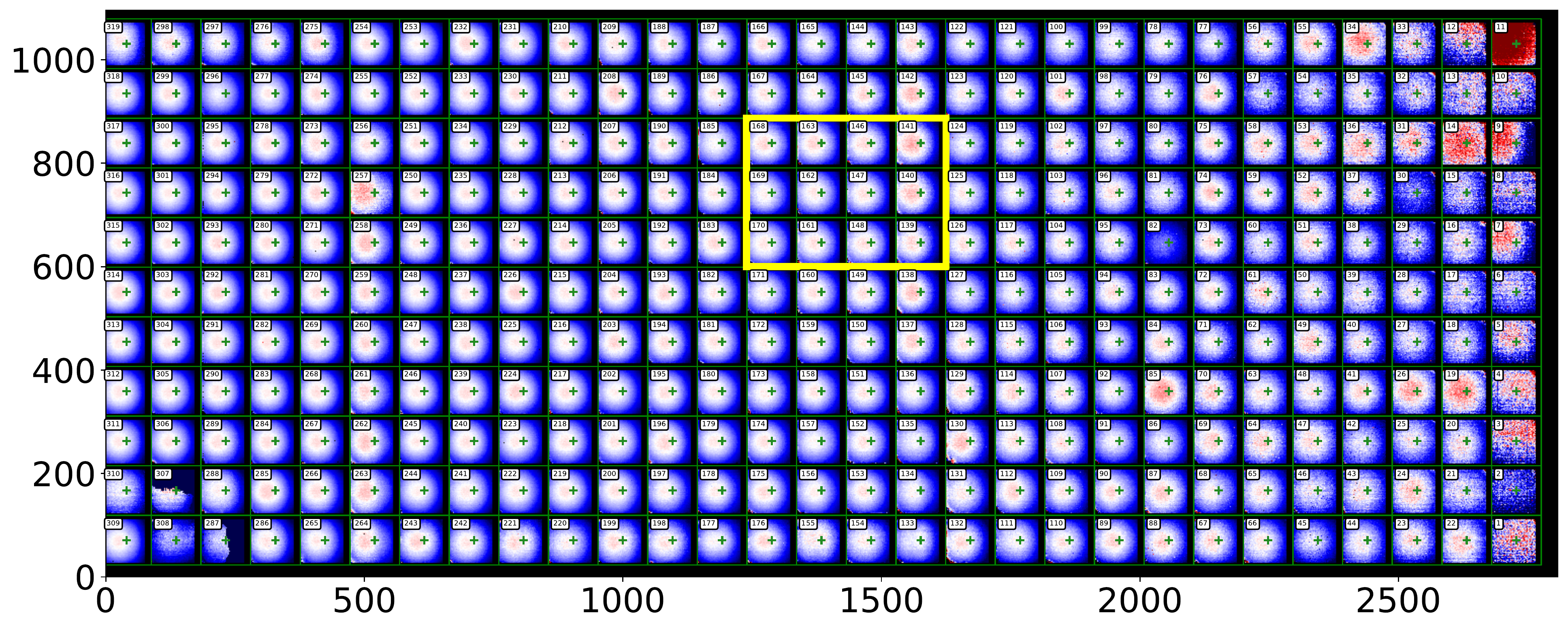}
				\caption{\Glsentryshort{opd} increase. Visualization of the entire focal plane array.}
				\label{fig:incidence_angle_fpa}
			\end{subfigure}
			\hspace*{0.06\linewidth}
			\\
			\begin{subfigure}[b]{0.45\linewidth}
				\centering
				\includegraphics[width=\linewidth]{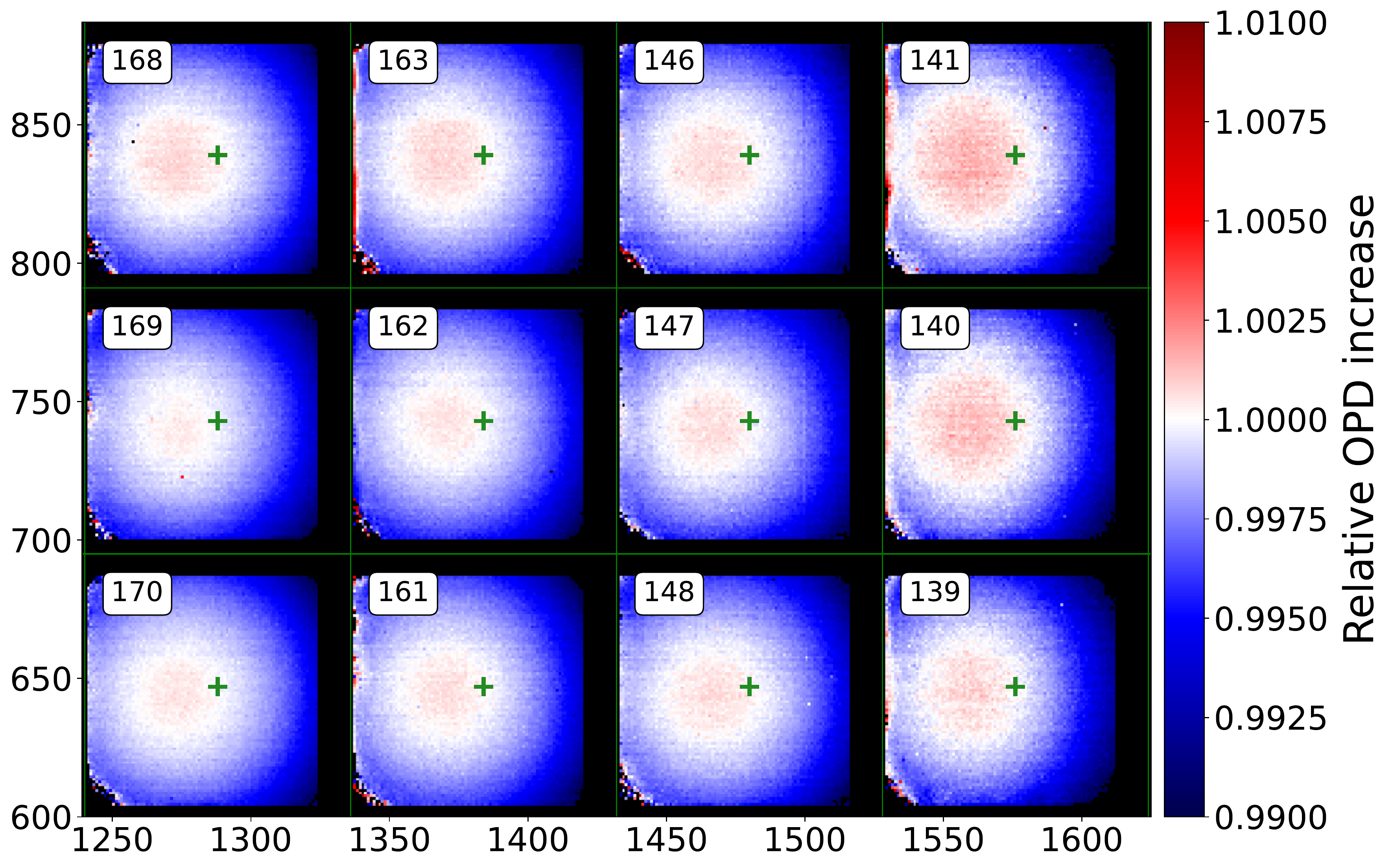}
				\caption{\Glsentryshort{opd} increase. Zoomed-in visualization.}
				\label{fig:incidence_angle_zoom_0}
			\end{subfigure}
			\begin{subfigure}[b]{0.45\linewidth}
				\centering
				\includegraphics[width=\linewidth]{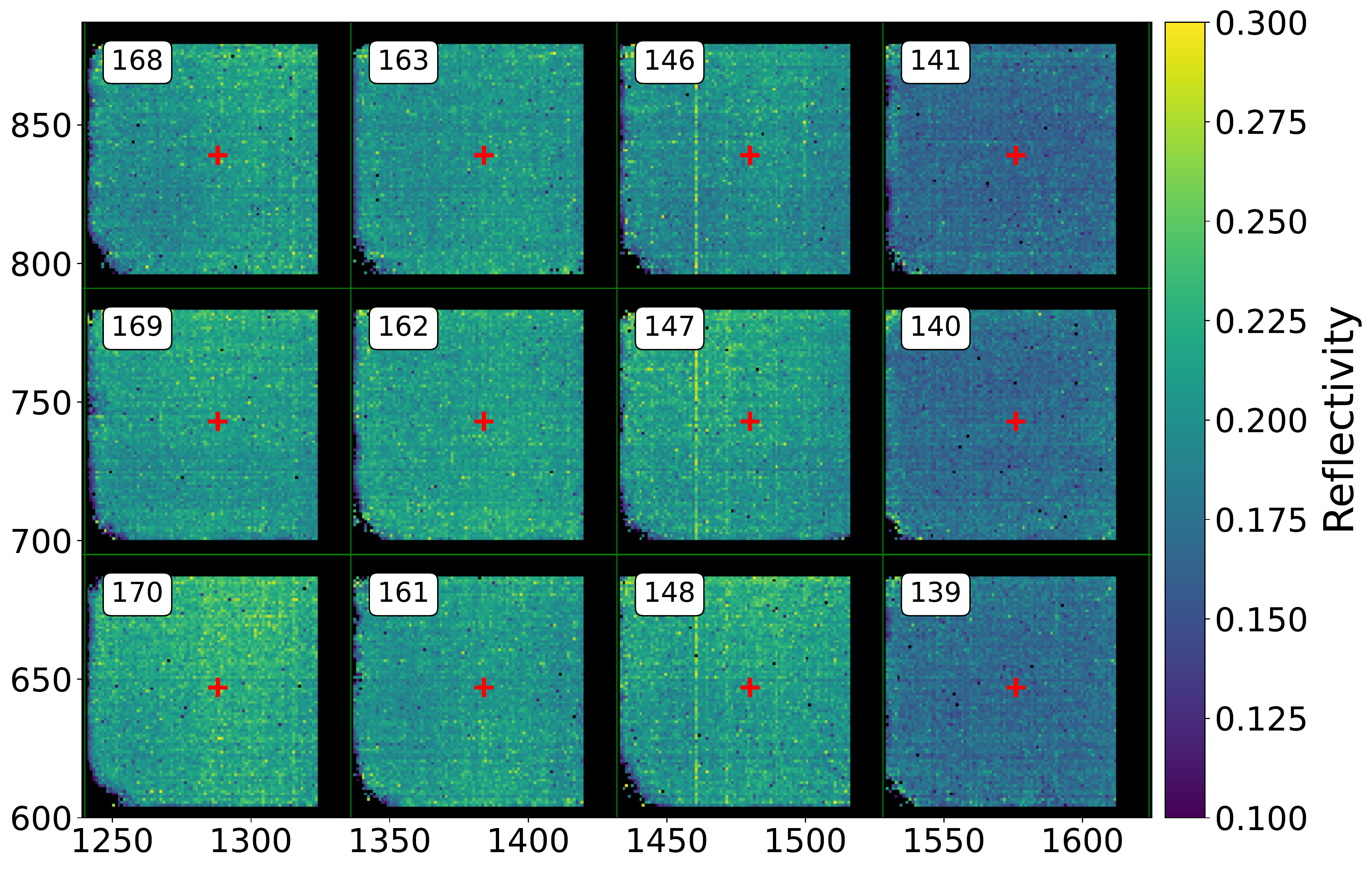}
				\caption{Reflectivity values in the zoomed area.}
				\label{fig:incidence_angle_zoom_1}
			\end{subfigure}
			\caption[Effect of incidence angle on the OPD]{
				Characterization of the \glsentryshort{p2} over the entire \glsentrylong{fov} of the device using the \glsentryshort{irca} method. (\subref{fig:incidence_angle_fpa}) Illustration of the relative increase of the estimated \glsentryshort{opd} compared to its value at the center of the corresponding subimage, indicated by a cross.
				Red and blue shades indicate values larger and smaller than the reference, respectively. Black pixels correspond to areas where the acquisition data was unavailable or where the \glsentryshort{lm} algorithm did not converge within 100 iterations.
				(\subref{fig:incidence_angle_zoom_0}) Zoomed-in visualization of the yellow framed area.
				(\subref{fig:incidence_angle_zoom_1}) Estimated reflectivity per pixel over the same zoomed area, averaged over the wavenumber range of the device.
			}
			\label{fig:incidence_angle}
		\end{figure*}

		We also employed the proposed algorithm for the spectral characterization of \glsentryshort{p2} at different angles of incidence, following the common practice in the literature~\cite{Zhan21a, Feng23, Yang22}.
		Specifically, we conducted a comprehensive characterization of the transmittance response for every pixel on the focal plane using the \glsentryshort{ml}+\glsentryshort{gn} variant of the proposed \glsentryshort{irca} method. The resulting estimation for the \glsentryshort{opd} is displayed in \figurename~\ref{fig:incidence_angle}, where the subimage area of the instrument corresponds to a \glsentrylong{fov} of around $\pm 5$ degrees.
		Our measurements indicate that the relative variation of the \gls{opd} with respect to its value in the central pixel is experimentally found to be more or less identical across different interferometers.
		The analysis shows a decreasing trend for the \gls{opd} which emanates radially from a central position. This is an expected result from eq.~\eqref{eq:opd_fabry_perot}, as the \gls{opd} decreases when the angle of incidence increases. The true optical axis is however shifted to the left with respect to the geometrical center that we have arbitrarily chosen as center of the subimage, as shown in \figurename~\ref{fig:incidence_angle_zoom_0}. On the contrary, the estimated values of the reflectivity preserve a certain flatness within their own subimage (\figurename~\ref{fig:incidence_angle_zoom_1}).
		The spatial analysis also allows for the detection of certain instrument defects, such as those of two of the subimages in bottom left area of \figurename~\ref{fig:incidence_angle_fpa}. In a significant portion of these subimages, the device exhibits behavior that significantly deviates from the expected one, suggesting the presence of defects in the reflective coating.

\section{Conclusion}

	In this paper, we presented a characterization procedure for multi-aperture imaging spectrometers based on \glsentrylong{fp} interferometers.
	We described the image formation model and we expressed its transmittance response in terms of a limited amount of parameters, following the formulation of Airy distribution and describing different regimes of finesse under the same framework. The proposed characterization algorithm exploits the two emerging wave model to cast the problem in the Fourier domain, where the \gls{ml} estimator for the \glspl{opd} is equivalent to a maximization of the periodogram. This result is then refined through nonlinear regression. Using the proposed multiple-step procedure in the algorithm allows for both robustness and precision in the final results.
	The estimated parameters can highlight manufacturing issues in an easily interpretable format. A proper characterization is extremely important for a proper recovery of the spectrum, which in the future could be optimized jointly with the spectral response of the devices in architectures where such parameters can be learned dynamically~\cite{Sun16, Yang20}.
	
\appendix

\section{Levenberg-Marqardt algorithm}
\label{ssec:lm}

	We aim to provide here an approachable explanation of the \gls{lm} algorithm, to build an intuition of what are the operations involved in the process.
	The algorithm can be seen as a trust region based approach for nonlinear regression. 
	The aim is to find an estimation $\hat{\bm{\beta}}$ of the parameters $\bm{\beta}=\{\beta_m\}_{\range{i}{1}{N_m}}$, in order for the samples $\{t_i(\bm{\beta})\}_{\range{i}{1}{N_a}}$ of an analytical function to fit a set of observation $\{y_i\}_{\range{i}{1}{N_a}}$.
	
	The algorithm addresses the problem by finding a sequence of iteratively more accurate solutions $\{\bm{\beta}^{[q]}\}_{q\ge0}$ from a given initialization $\bm{\beta}^{[0]}$, using the following update rule:
	\begin{equation}
		\bm{\beta}^{[q]}=\arg\min_{\bm{\beta}}\sum_{i=1}^{N_a}\left(t_i(\bm{\beta}^{[q-1]})+\sum_{m=1}^{N_m}j_{im}\left(\beta_m-\beta_m^{[q-1]}\right)-w_i\right)^2+\lambda\sum_{m=1}^{N_m}\beta_m^2\,.
		\label{eq:cost_lm}
	\end{equation}
	where $\lambda\ge0$ denotes an user defined dampening parameter.
	In the above function, $t_i(\bm{\beta})\approx t_i(\bm{\beta}^{[q-1]})+\sum_{m=0}^{N_m-1}j^{[q-1]}_{im}\left(\beta_m-\beta_m^{[q-1]}\right)$ represents a truncated Taylor expansion of $t(\bm{\beta})$ around the current estimation $\bm{\beta}^{[q-1]}$. In this representation, the terms $j_{im}$ denote the coefficients of the Jacobian matrix $\mathbf{J}\in\mathbb{R}^{N_p\times N_m}$, which are defined as:
	\begin{equation}
		j_{im}= \left.\frac{\partial t_i(\bm{\beta})}{\partial \beta_m}\right\rvert_{\bm{\beta}=\bm{\beta}^{^{[q-1]}}}\,,\;\;\;\;\;\;\forall\range{i}{1}{N_a},\,\forall\range{m}{1}{N_m}\,.
	\end{equation}
	
	Eq.~\ref{eq:cost_lm} admits as analytical solution:
	\begin{equation}
		\bm{\beta}^{[q]}=\bm{\beta}^{[q-1]}+\left(\mathbf{J}\T\mathbf{J}+\lambda\mathbf{I}\right)^{-1}\mathbf{J}\T\mathbf{e}\,,
	\end{equation}
	where $\mathbf{I}$ denotes an identity matrix and the vector $\mathbf{e}$, whose $i$-th coefficient is defined as $e_i:=t_i(\bm{\beta}^{[q-1]})-w_i$, denotes the current estimation residual. When a certain convergence condition is verified (e.g. after a given number of iterations), the result of the last update is chosen as the desired estimation $\hat{\bm{\beta}}$. Additional implementation details, e.g. to define a criterion to assign the value of $\lambda$, to simplify the evaluation of $\mathbf{J}$, and to evaluate the stopping conditions on the iterations are given in the related paper~\cite{More78}.

\section{Backmatter}

\begin{backmatter}
	For illustration, authors Yann Ferrec, and Etienne Le Coarer, are represented below as YF, and ELC.
	
	\bmsection{Funding}
	AuRA region and FEDER (ImSPOC-UV: convention FEDER n. RA0022348); Agence Nationale de la Recherche (FuMultiSPOC: ANR-20-ASTR-0006); Agence Nationale de la Recherche (LabEx FOCUS: ANR-11-LABX-0013).
	
	\bmsection{Disclosures}
	\noindent YF, ELC: ONERA, UGA - U.S. Patent:  10,677,650 B2 (P).
	
	\bmsection{Data Availability} Data underlying the results presented in this paper are available in Ref.~\cite{web_IRCA}.

\end{backmatter}

\bibliography{biblio_oe.bib}

\end{document}